\documentclass[aps,prl,superscriptaddress,twocolumn]{revtex4-1} 

\usepackage{amsfonts}
\usepackage{subfigure}
\usepackage{amsmath}
\usepackage{txfonts}
\usepackage{amssymb}
\usepackage{amsbsy} 
\usepackage{epsfig}
\usepackage{color}

\def\be{\begin{equation}} \def\ee{\end{equation}}
\def\bea{\begin{eqnarray}} \def\eea{\end{eqnarray}}

\def\nn{\nonumber}

\def\bk{{\bf k}}

\def\bD{{\bf D}}

\def\be{{\bf e}}
\def\bd{{\bf d}}

\def\la{\langle}
\def\ra{\rangle}

\def\rw{\rightarrow}

\begin{document}

\title{Topological defects in Floquet systems: Anomalous chiral modes and topological invariant}

\author{Ren Bi} \affiliation{ Institute for
Advanced Study, Tsinghua University, Beijing, 100084,  China}

\author{Zhongbo Yan} \affiliation{ Institute for
Advanced Study, Tsinghua University, Beijing, 100084,  China}

\author{Ling Lu}
\affiliation{ Institute of Physics, Chinese Academy of Sciences/Beijing National Laboratory for Condensed Matter Physics, Beijing 100190, China }

\author{Zhong Wang}
\altaffiliation{  wangzhongemail@tsinghua.edu.cn} \affiliation{ Institute for
Advanced Study, Tsinghua University, Beijing, 100084,  China}

\affiliation{Collaborative Innovation Center of Quantum Matter, Beijing, 100871, China }


\begin{abstract}

Backscattering-immune chiral modes arise along certain line defects in three-dimensional materials. In this paper, we study Floquet chiral modes along Floquet topological defects, namely, the defects come entirely from spatial modulations of periodic driving. We define a precise topological invariant that counts the number of Floquet chiral modes, which is expressed as an integral on a five-dimensional torus parameterized by $(k_x,k_y,k_z,\theta,t)$. This work demonstrates the possibility of creating chiral modes in three-dimensional bulk materials by modulated driving. We hope that it will stimulate further studies of Floquet topological defects.

\end{abstract}

\maketitle

Chiral edge states\cite{laughlin1981,halperin1982,wen1990} are hallmarks of quantum (anomalous) Hall effects\cite{klitzing1980,Haldane1988,chang2013experimental,yu2010a}. The number of chiral edge modes is determined by the first Chern number\cite{thouless1982,niu1985,Hatsugai1993} of the occupied bands of the two-dimensional(2D) systems, which is a best example of bulk-boundary correspondence in topological phases\cite{hasan2010,qi2011,Chiu2015RMP,Bansil2016,chen2013symmetry,wen2004}. Due to complete absence of backscattering channel, transport by chiral modes is dissipationless (as exemplified by the vanishing longitudinal resistivity\cite{klitzing1980}), which is potentially important in future low-power electronics.

Time-dependent external fields, such as monochromatic lasers, offer highly controllable and tunable tools for creating topological band structures, enlarging the experimental frontiers of topological materials. Recently, considerable progresses have been made, both theoretically\cite{Torres2014,Dahlhaus2011,Gomez2013,
Zhou2011Optical,Delplace2013,wang2014floquet,DAlessio2014,
Seetharam2015,titum2016,Goldman2015,Wang2016Network,Hubener2016, Else2016,Mori2016,Lazarides2015,Khemani2016, Zhou2016Floquet,Thakurathi2013,wang2011Long,Usaj2014} and experimentally\cite{wang2013observation,
mahmood2016selective,rechtsman2013photonic,gao2016probing,Stehlik2016}, in understanding periodically driven (Floquet) systems, particularly in connection with topological phases\cite{Oka2009,lindner2011floquet,Kitagawa2011,Inoue2010,
Gu2011,Kitagawa2010a,Kitagawa2010b, Jiang2011,rudner2013anomalous,Carpentier2015,kitagawa2012observation,Karzig2015,fang2012realizing, Dora2012,cayssol2013floquet,Chan2016hall,yan2016tunable,
Narayan2016,Chan2016type,Roy2016periodic,qu2016}.
Remarkably, chiral edge state can exist even if the first Chern number of every bulk band vanishes\cite{Kitagawa2010b,rudner2013anomalous,Hu2015,Leykam2016,Mukherjee2016}. This phenomenon is closely related to the absence of band bottom, which is a distinctive feature of Floquet systems\cite{rudner2013anomalous}.

Interestingly, certain line defects in 3D crystals also host chiral modes\cite{callan1985,witten1985superconducting,wang2013a,Bi2015, Schuster2016cable,Roy2015magnetic,You2016}\footnote{Helical defect modes were also studied\cite{ran2009one,Slager2014,zhang2015}.}, which are protected by the second Chern number\cite{teo2010,qi2008,fiber}. Experimental realization is lacking so far, because it is challenging to create and manipulate line defect in a controllable manner, therefore, it is worthwhile to study \emph{Floquet defects}, which can be created by spatial modulation of the driving field\cite{Katan2013modulated}, without the need of preexisting static defect. The Floquet chiral channels have the advantages that they can be opened or closed,  and their spatial locations can readily be tuned, by external driving. Several intriguing questions arise in this direction. How to determine the number of Floquet chiral modes? How to create them?
In this paper, we construct a topological invariant expressed in terms of the evolution operator (inspired by Refs.\cite{rudner2013anomalous,Carpentier2015}). It is defined as an integral on the 5D torus parameterized by $(k_x,k_y,k_z,\theta,t)$ [Eq.(\ref{W})], combining three types of coordinates: \emph{momentum, space, time}. We then construct a concrete lattice model, showing that appropriate modulations indeed generate Floquet chiral modes, in agreement with the prediction from topological invariant.

\begin{figure}
\includegraphics[width=8cm, height=4.2cm]{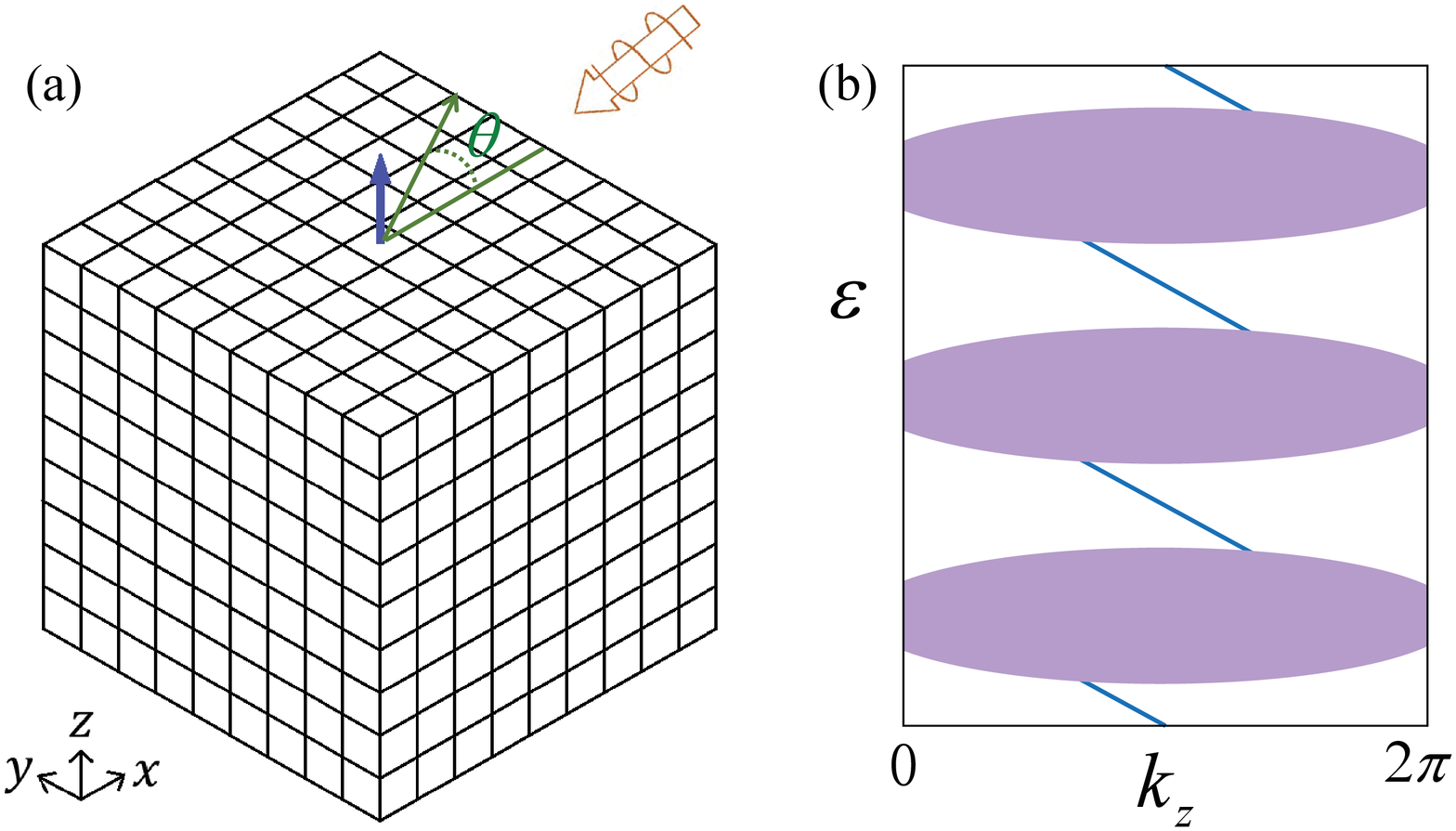}
\caption{ Sketch. (a) The time-periodic driving is spatially modulated as a function of the polar angle $\theta$, creating a Floquet line defect along $r=0$ [$r\equiv \sqrt{(x-x_0)^2+(y-y_0)^2},\,\tan\theta\equiv (y-y_0)/(x-x_0)$; $(x_0,y_0)$ is the location of defect]. The blue arrow stands for the Floquet chiral modes inside the bulk energy gap. (b) Sketch of the quasienergy bands, with shadow region representing the bulk bands, connected by the Floquet chiral modes (blue lines). }  \label{sketch}
\end{figure}

{\it Topological invariant.--}We will define a topological invariant that counts the number of Floquet chiral modes along a line defect. Let us take the cylindrical coordinates $ (r, \theta,z)$ with the defect located along $r=0$, so that the Hamiltonian varies with $\theta$. The topological information can be read from the regions distant from the defect. Sufficiently far away from the defect, the spatial variation of Hamiltonian is slow, which allows us to define the crystal momentum $\bk=(k_x,k_y,k_z)$\cite{teo2010}. The Bloch Hamiltonian $H(\bk,\theta,t)$ is an $s\times s$ matrix
($s$ is the number of bands), which is periodic in time: $H(\bk,\theta,t)=H(\bk,\theta,t+T)$, with $T=2\pi/\omega$.  We can define the time evolution operator $U(\bk,\theta,t)= \mathcal{T}\exp\left(-i\int_0^t dt' H(\bk,\theta,t')\right)$, where $\mathcal{T}$ denotes time ordering. The full-period evolution $U(\bk,\theta,T)$ can be diagonalized as \bea U(\bk,\theta,T)=\sum_{n=1}^s\lambda_n|\psi_n \ra\la\psi_n|, \eea
and an effective Hamiltonian $H_\varepsilon^{\rm eff}$ can be defined: \bea H_\varepsilon^{\rm eff}(\bk,\theta)=\frac{i}{T}\sum_n\ln_\varepsilon(\lambda_n)|\psi_n\ra\la\psi_n|, \label{ln} \eea where $\ln_\varepsilon$ is the logarithm with branch cut at $e^{-i\varepsilon T}$, namely $\log e^{-i\varepsilon T+i0^+}=\log e^{-i\varepsilon T+i0^-}-2\pi i=-i\varepsilon T$\cite{rudner2013anomalous,Carpentier2015}.
It is apparent that $U(\bk,\theta,T)=\exp[-iH_\varepsilon^{\rm eff}(\bk,\theta)T]$. To have smooth dependence of $H^{\rm eff}_\varepsilon$ on $\bk$ and $\theta$, $e^{-i\varepsilon T}$ must lie in an eigenvalue gap of $U(\bk,\theta,T)$. The coefficients $\varepsilon_n=(i/T)\ln_\varepsilon(\lambda_n)$ in Eq.(\ref{ln}) are known as quasienergies.

Now we construct a periodic version of $U$\cite{rudner2013anomalous,Carpentier2015}: \bea U_\varepsilon(\bk,\theta,t)=U(\bk,\theta,t)\exp[i H^{\rm eff}_\varepsilon(\bk,\theta)t], \label{U-epsilon} \eea which satisfies $U_\varepsilon(\bk,\theta,T)=I=U_\varepsilon(\bk,\theta,0)$. This property enables us to define the integer topological invariant
\bea W(\varepsilon) =  &&\frac{i}{480\pi^3}\int dt d\theta  d^3k \,{\rm Tr} [\epsilon^{\mu\nu\rho\sigma\tau} (U_\varepsilon^{-1}\partial_\mu U_\varepsilon )( U_\varepsilon^{-1}\partial_\nu U_\varepsilon ) \nn \\ && \times (U_\varepsilon^{-1}\partial_\rho U_\varepsilon ) (U_\varepsilon^{-1}\partial_\sigma U_\varepsilon)(   U_\varepsilon^{-1}\partial_\tau U_\varepsilon)], \label{W} \eea where the integrating range is $[0,T]\times[0,2\pi]\times{\rm BZ}$ (BZ=Brillouin zone), $\mu,\nu,\rho,\sigma,\tau=k_x,k_y,k_z,\theta,t$, and  $\epsilon^{\mu\nu\rho\sigma\tau}=\pm 1$ is the Levi-Civita symbol. Given the evolution operator in the 5D parameter space $(k_x,k_y,k_z,\theta,t)$, Eq.(\ref{W}) seems to be the only natural topological invariant. The normalization factor in Eq.(\ref{W}) ensures that $W$ is integer-valued\cite{bott1978some,witten1983,wang2010b,wang2012a}.
As a test, we can show\cite{supplemental} that $W$ reduces in static systems to the second Chern number\cite{teo2010,qi2008}, which is known to count the number of chiral modes along static defects\cite{teo2010,fiber}.

Given two quasienergy gaps $0\leq\varepsilon<\varepsilon'<\omega$, the difference in the branch cut of logarithm causes $H^{\rm eff}_{\varepsilon'}(\bk,\theta)-H^{\rm eff}_\varepsilon(\bk,\theta)=\omega P_{\varepsilon,\varepsilon'}$, where $P_{\varepsilon,\varepsilon'}=\sum'_{n}|\psi_n\ra\la\psi_n|$ is a projection operator, $\sum'_n$ denoting summation for $\varepsilon<{\rm arg}(1/\lambda_n)/T<\varepsilon'$. One can define the second Chern number in this subspace, $C_2(\varepsilon,\varepsilon')= (-1/8\pi^2)\int d\theta d^3k\,{\rm Tr}[\epsilon^{ijkl}P_{\varepsilon,\varepsilon'} \partial_iP_{\varepsilon,\varepsilon'} \partial_jP_{\varepsilon,\varepsilon'} P_{\varepsilon,\varepsilon'} \partial_kP_{\varepsilon,\varepsilon'}\partial_lP_{\varepsilon,\varepsilon'}]$, in which $i,j,k,l=k_x,k_y,k_z,\theta$. This projection-operator expression is equivalent to the Berry-curvature expression\cite{qi2008}. From the observation $U^{-1}_{\varepsilon}U_{\varepsilon'}=\exp(i\omega t P_{\varepsilon,\varepsilon'})$, one can show that\cite{supplemental} \bea W(\varepsilon')-W(\varepsilon)=C_2(\varepsilon,\varepsilon'), \label{WW} \eea  therefore, the Chern number measures the difference between the numbers of chiral modes above and below the band, which, due to the absence of band-bottom, cannot fully determine the number of chiral modes in each gap.

{\it Model.--}Eq.(\ref{W}) provides clues to model design. For instance, if $U$ matrix is $2\times 2$ (namely two-band), then $W=0$; thus we consider four-band models. Before investigating spatially modulated driving, we study the homogeneous system first. The Hamiltonian reads
\bea H(\bk,t) = H_0(\bk) + H_d(t), \eea
where the first part $H_0$ describes a Dirac semimetal:
\bea H_0(\bk)= && (2t_x\sin k_x\sigma_x + 2t_y\sin k_y\sigma_y + 2t_z\sin k_z\sigma_z)\otimes\tau_z  \nn \\ && + m(\bk)\sigma_0\otimes\tau_x, \label{static} \eea in which $\sigma_{x,y,z}$ and $\tau_{x,y,z}$ are Pauli matrices ($\sigma_0=\tau_0=I$), $m(\bk)=B_0-B_1\sum_{i=x,y,z}\cos k_i-B_2\sum_{i\neq j}\cos k_i\cos k_j$, with parameters $t_{x,y,z}=B_1=1, B_2=0.1, B_0=3.6$. Near the Dirac point $(0,0,0)$, $H_0(\bk)\approx\sum_i 2t_i k_i\sigma_i\otimes\tau_z$. The second part $H_d$ is a driving: \bea H_d(t) = 2 D\cos(\omega t) \sigma_0\otimes (\tau_x\cos\alpha+\tau_y\sin\alpha), \label{Hd}  \eea with $\alpha$ to be specified shortly. Eq.(\ref{static}) can be written compactly as $H_0={\bf d}\cdot{\bf \Gamma}\equiv\sum_{\mu=1}^5 d_\mu \Gamma_\mu$, with $d_{1,2,3}=2t_{x,y,z}\sin k_{x,y,z}$, $d_4=m(\bk)$, $d_5=0$, $\Gamma_{1,2,3}=\sigma_{x,y,z}\otimes\tau_z$, $\Gamma_4=\sigma_0\otimes\tau_x$, and $\Gamma_5=\sigma_0\otimes\tau_y$. The bands of $H_0$ are $E_{\pm}(\bk)=\pm d(\bf{k})$ ($d\equiv|{\bd}|$).

Physically, we can regard $\sigma_z=\pm 1$ as spin states, and $\tau_z=\pm 1$ as two orbitals. Suppose that $\tau_z=\pm 1$ orbitals have adjacent orbital-angular-momentum quantum numbers, say $m_z=0,1$, respectively, then $H_d(t)$ describes the electric-dipole coupling to an alternating electric field in the $(\cos\alpha,\sin\alpha,0)$ direction, therefore, $H_d(t)$ can be provided by a linearly-polarized laser beam with frequency $\omega$.

\begin{figure}
\subfigure{\includegraphics[width=4.0cm, height=3.8cm]{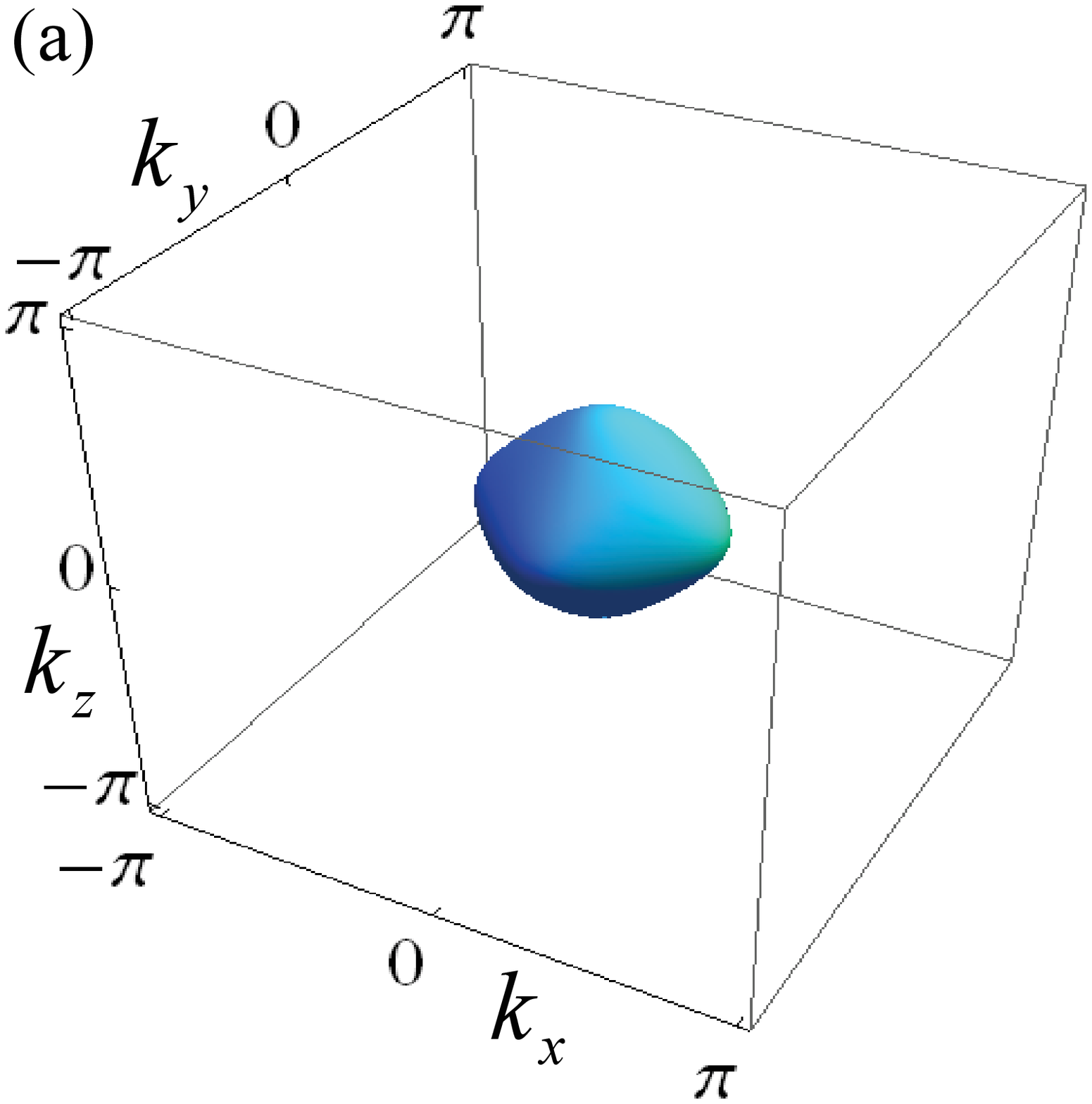}}
\subfigure{\includegraphics[width=4.4cm, height=3.8cm]{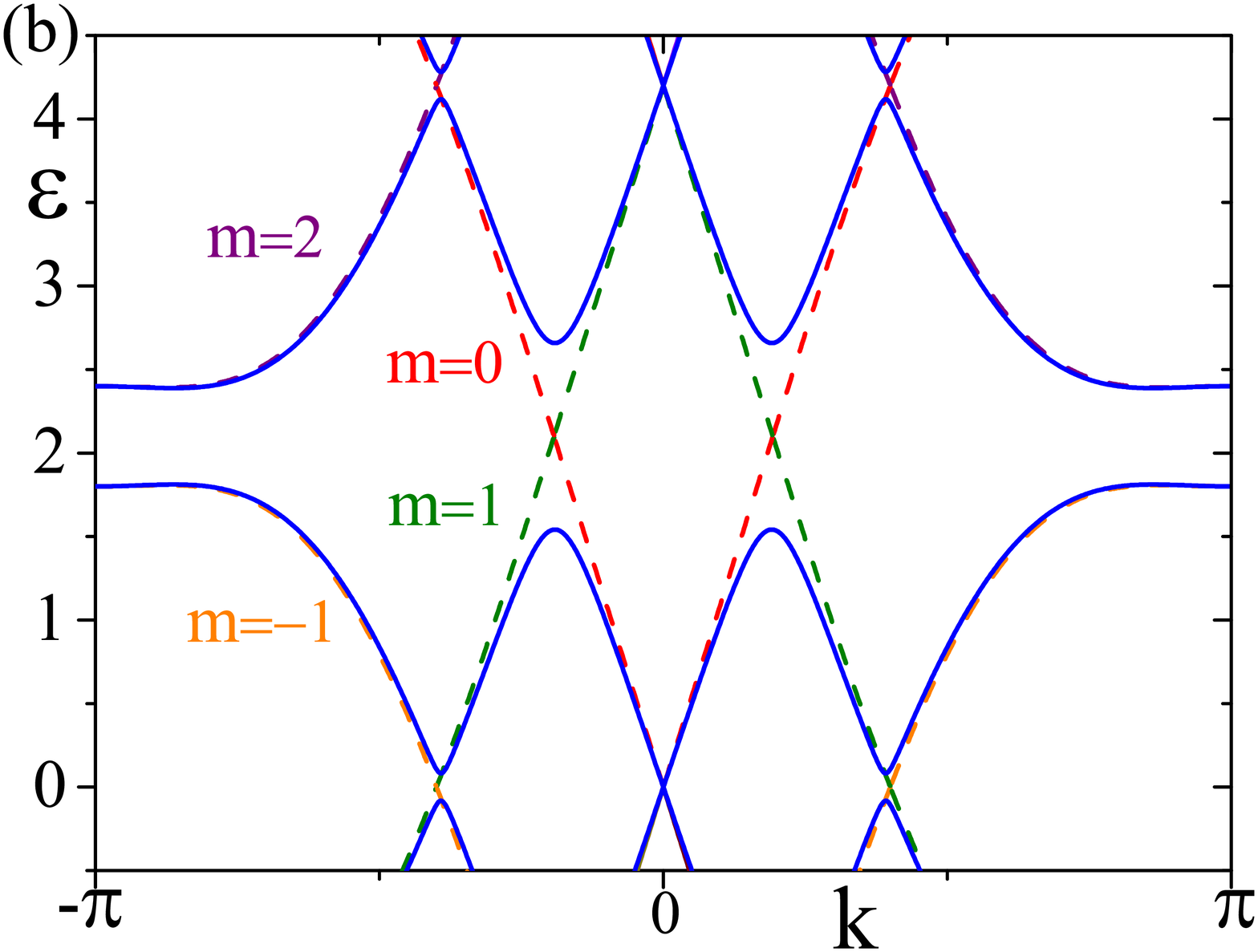}}
\caption{(a) The resonance surface $d(\bk)=\omega/2$, where the $m=0$ and $m=1$ Floquet bands cross each other (when $D=0$). (b) Floquet bands along 111 direction ($k_x=k_y=k_z=k$) for $D=0$ (dashed curves, with Floquet index $m$ marked), and $D=0.6$ (solid curves).  All bands are doubly degenerate.  } \label{gap}
\end{figure}

We shall calculate the quasienergy bands $\varepsilon_n(\bk)$ in frequency domain.
Employing the Fourier expansion\cite{Oka2009,Kitagawa2011,rudner2013anomalous}, $|\psi_{n}(\bk,t)\ra=e^{-i\varepsilon_n(\bk)t}\sum_{m=-\infty}^{\infty} |\phi_n^{(m)}(\bk)\ra  e^{im\omega t}$, the Schr\"odinger equation $i\partial_t |\psi_n(\bk,t)\ra= H (\bk,t)|\psi_n(\bk,t)\ra$ becomes \bea\sum_{m'} \mathcal{H}_{mm'}(\bk) |\phi^{(m')}_n (\bk) \ra =\varepsilon_n (\bk) |\phi^{(m)}_n(\bk)\ra, \label{freq} \eea in which $\mathcal{H}_{mm'}=m\omega\delta_{mm'}{\bf 1}+H_{m-m'}$, with $H_m=\frac{1}{T}\int_0^Tdt e^{-im\omega t}H(t)$. The Floquet Hamiltonian in Eq.(\ref{freq}) is an infinite-rank matrix: \bea \mathcal{H}(\bk)=\left(
                                \begin{array}{ccccc}
                               \cdots &&&& \\
                                    & H_0+\omega & H_1 & H_2 &  \\
                                    & H_{-1} & H_0 & H_1 &  \\
                                    & H_{-2} & H_{-1} & H_0-\omega & \\
                                    &&&& \cdots \\
                                \end{array}
                                \right). \label{fH} \eea
The spectrum of Eq.(\ref{fH}) is mathematically equivalent to the Wannier-Stark ladder\cite{emin1987existence}, whose eigenstates are localized in $m$, namely, each eigenfunction decays as $\exp(-|m-m_0|\omega/\Lambda)$ for a certain $m_0$ ($\Lambda$ is the system's typical energy scale).
Therefore, we can truncate $\mathcal{H}(\bk)$ to $\mathcal{H}^{(N)}(\bk)$, which contains $N\times N$ blocks, $H_0$ being the central block.  As long as $N\gg\Lambda/\omega$, the truncation errors for eigenfunctions not close to the upper and lower truncation edges (approximately at $\pm N\omega/2$) are exponentially small and thus negligible.

In our model, $H_{\pm 1}= D\sigma_0\otimes(\tau_x\cos\alpha + \tau_y\sin\alpha)$ and $H_{\pm2,\pm 3,\cdots}=0$. When $D=0$, the Floquet bands are given by $E_\pm(\bk)+m\omega$. Adjacent Floquet bands cross at the resonance surface (Fig.\ref{gap}a) defined by $d(\bk)=\omega/2$, namely $E_-(\bk)+\omega=E_+(\bk)$. Nonzero $D$ hybridizes adjacent Floquet bands, say $m=0$ and $m=1$, generating a quasienergy gap near $\varepsilon=\omega/2$. Hereafter we take $\omega=4.2$ and $D=0.6$ for concreteness. Floquet bands for $\alpha=0$ are shown in Fig.\ref{gap}b. Other values of $\alpha$ give qualitatively similar bands.

{\it Floquet chiral modes.--}Eq.(\ref{W}) implies that suitable spatial modulations of the driving $H_d$ (with $H_0$ unchanged) can generate Floquet chiral modes. To this end, we take in Eq.(\ref{Hd}) $\alpha=n\theta$ ($n$ is a nonzero integer; $\theta$ is the polar angle), thus \bea H_{\pm 1}= D\sigma_0\otimes[\cos(n\theta)\tau_x + \sin(n\theta)\tau_y],\eea creating a Floquet line defect at $r=0$ (Fig.\ref{sketch}a). Taking the previous physical interpretation of the model, these defects can be generated by cylindrical vector beams of laser\cite{zhan2009cylindrical}, in which the spatial modulation in polarization takes exactly the desired form.

\begin{figure}
\subfigure{\includegraphics[width=8.5cm,height=4.6cm]{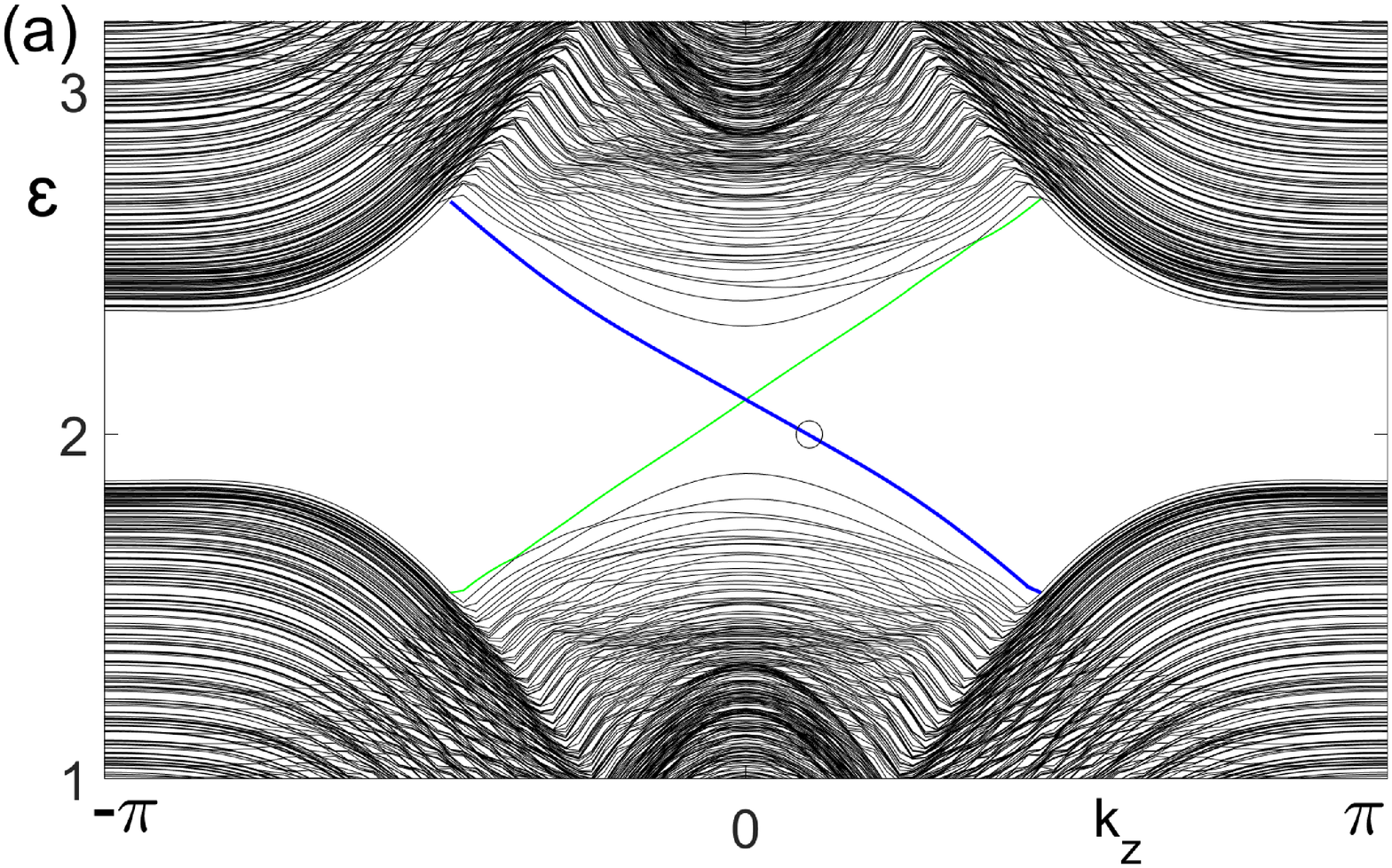}}
\subfigure{\includegraphics[width=4cm, height=3.3cm]{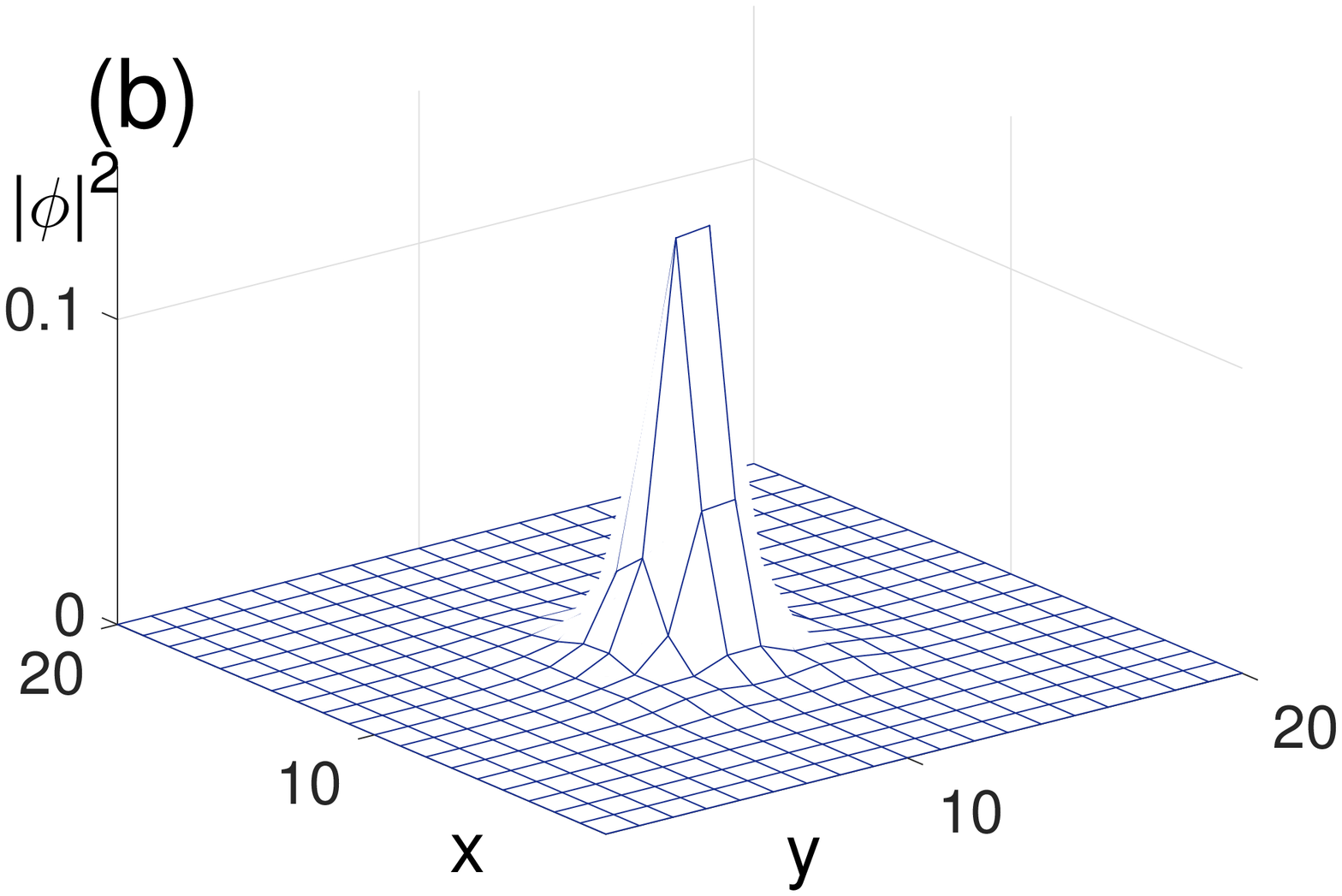}}
\subfigure{\includegraphics[width=4cm, height=3.3cm]{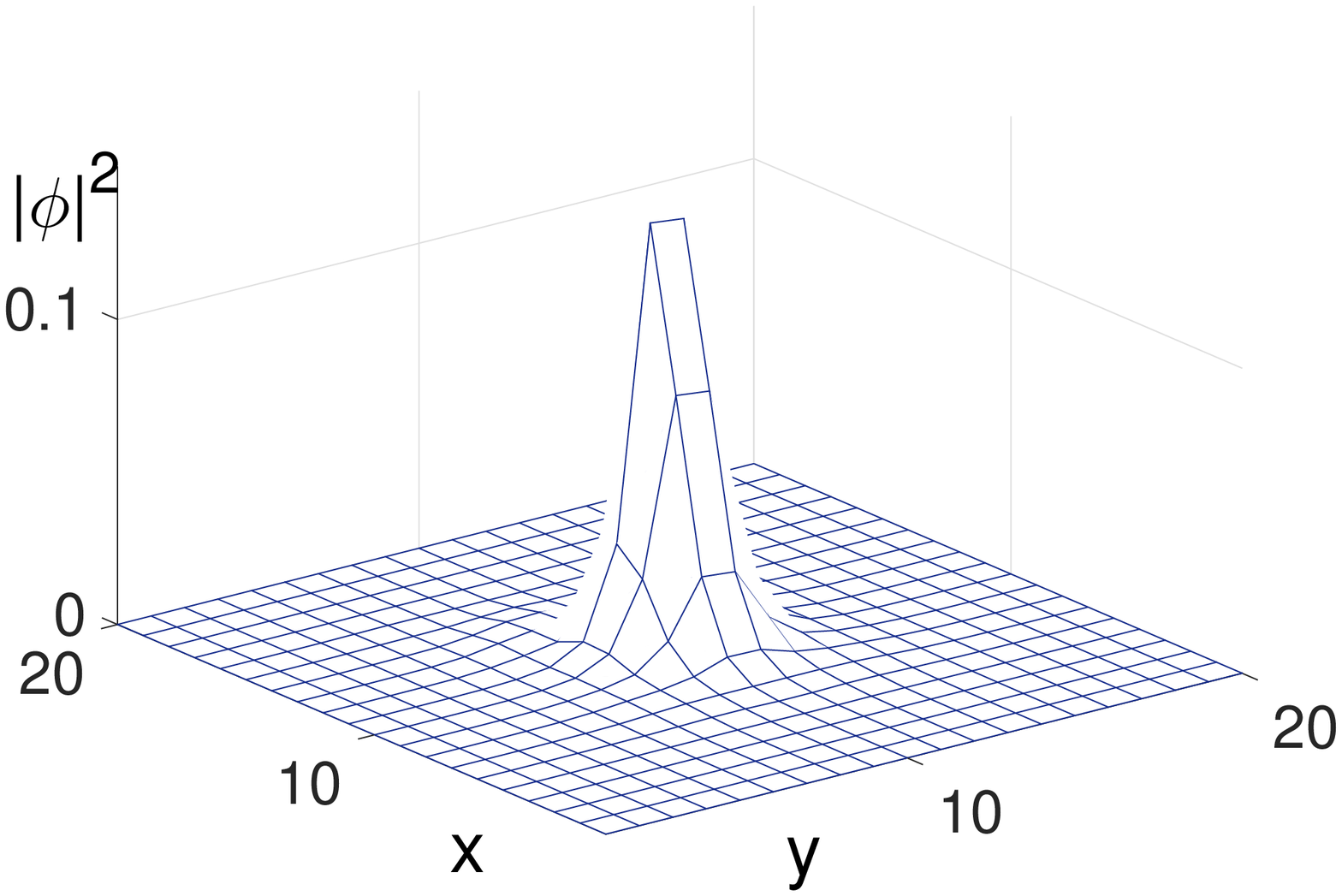}}
\subfigure{\includegraphics[width=8.5cm, height=4.6cm]{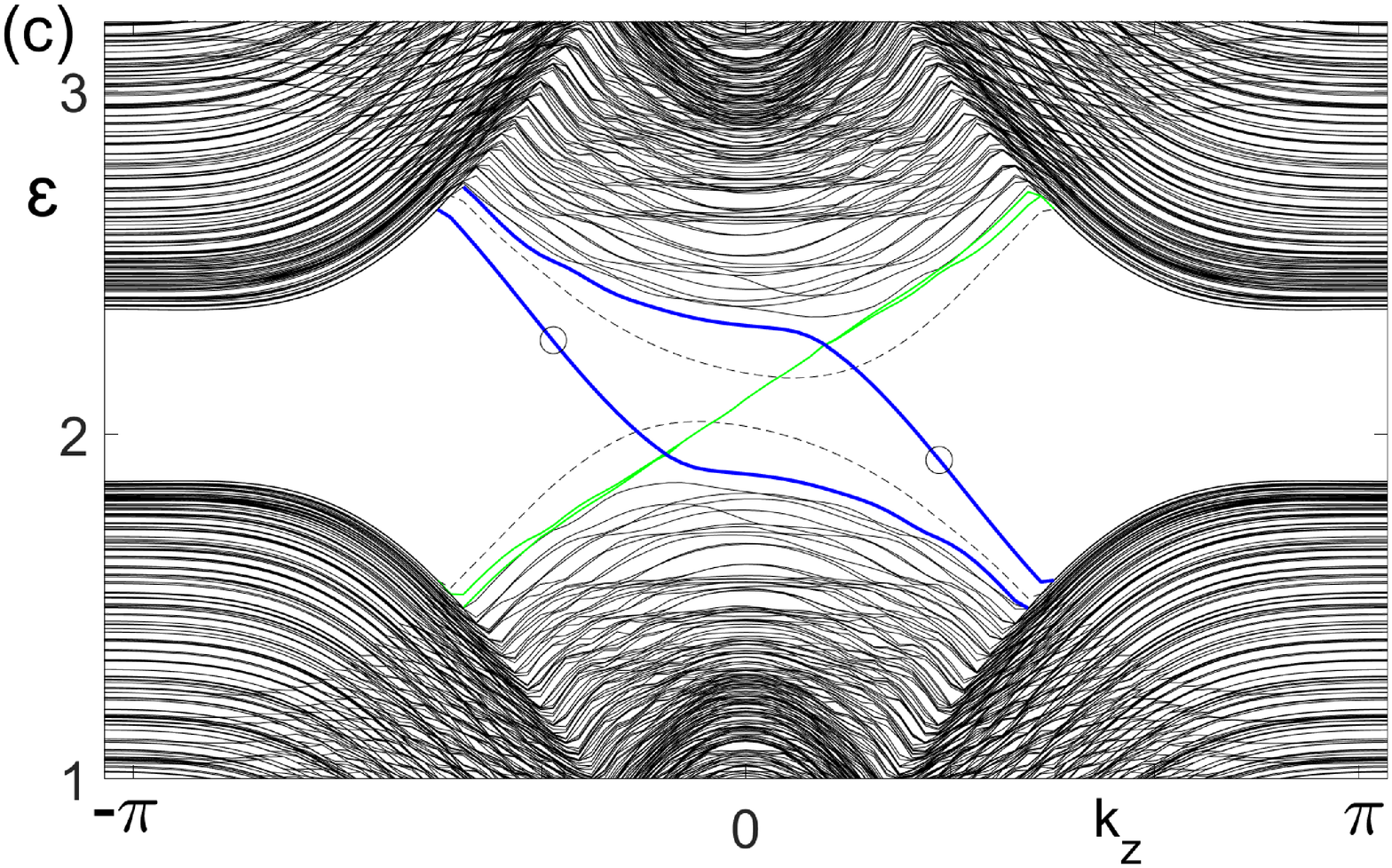}}
\subfigure{\includegraphics[width=4cm, height=3cm]{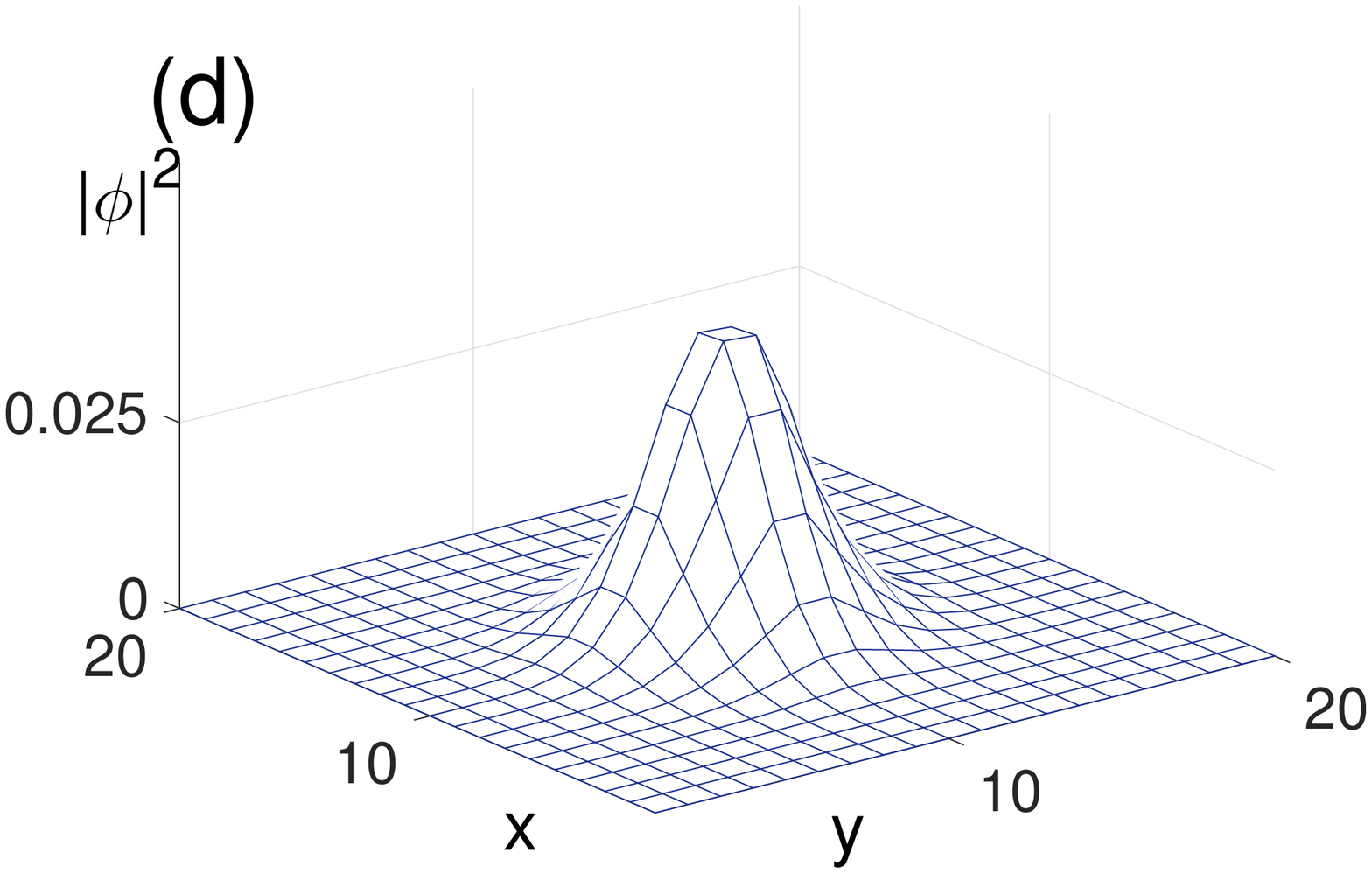}}
\subfigure{\includegraphics[width=4cm, height=3cm]{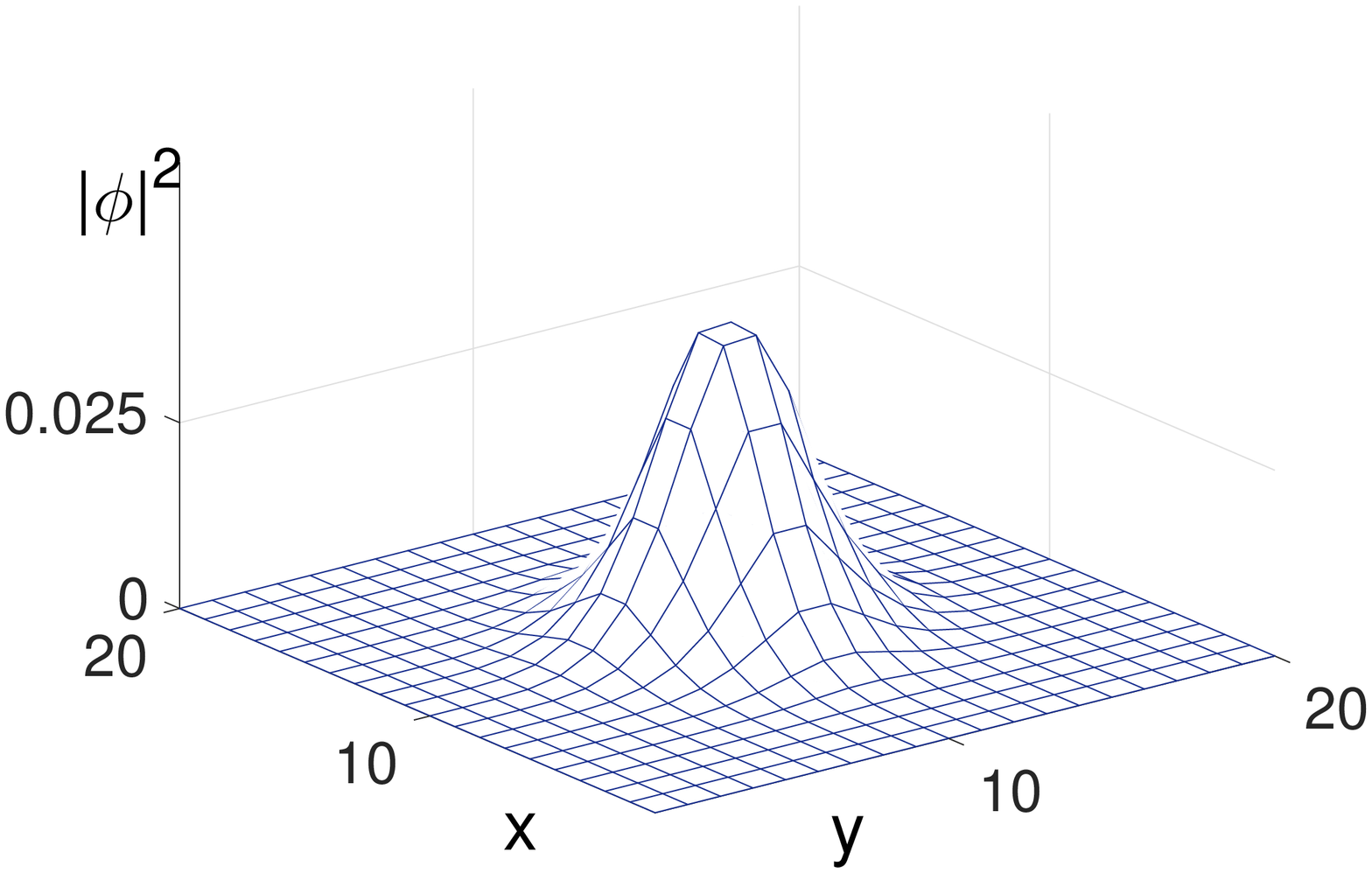}}
\subfigure{\includegraphics[width=4cm, height=3cm]{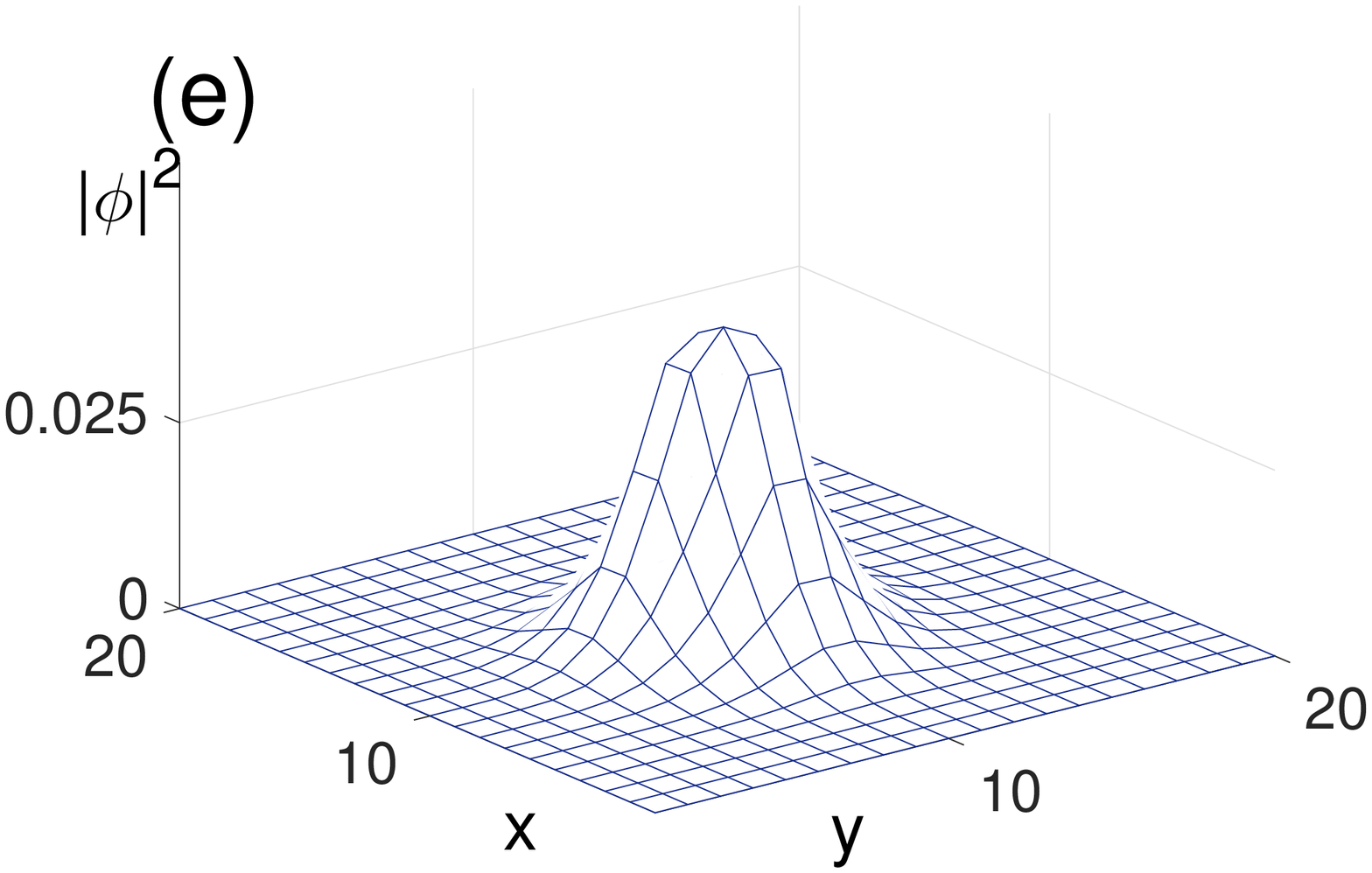}}
\subfigure{\includegraphics[width=4cm, height=3cm]{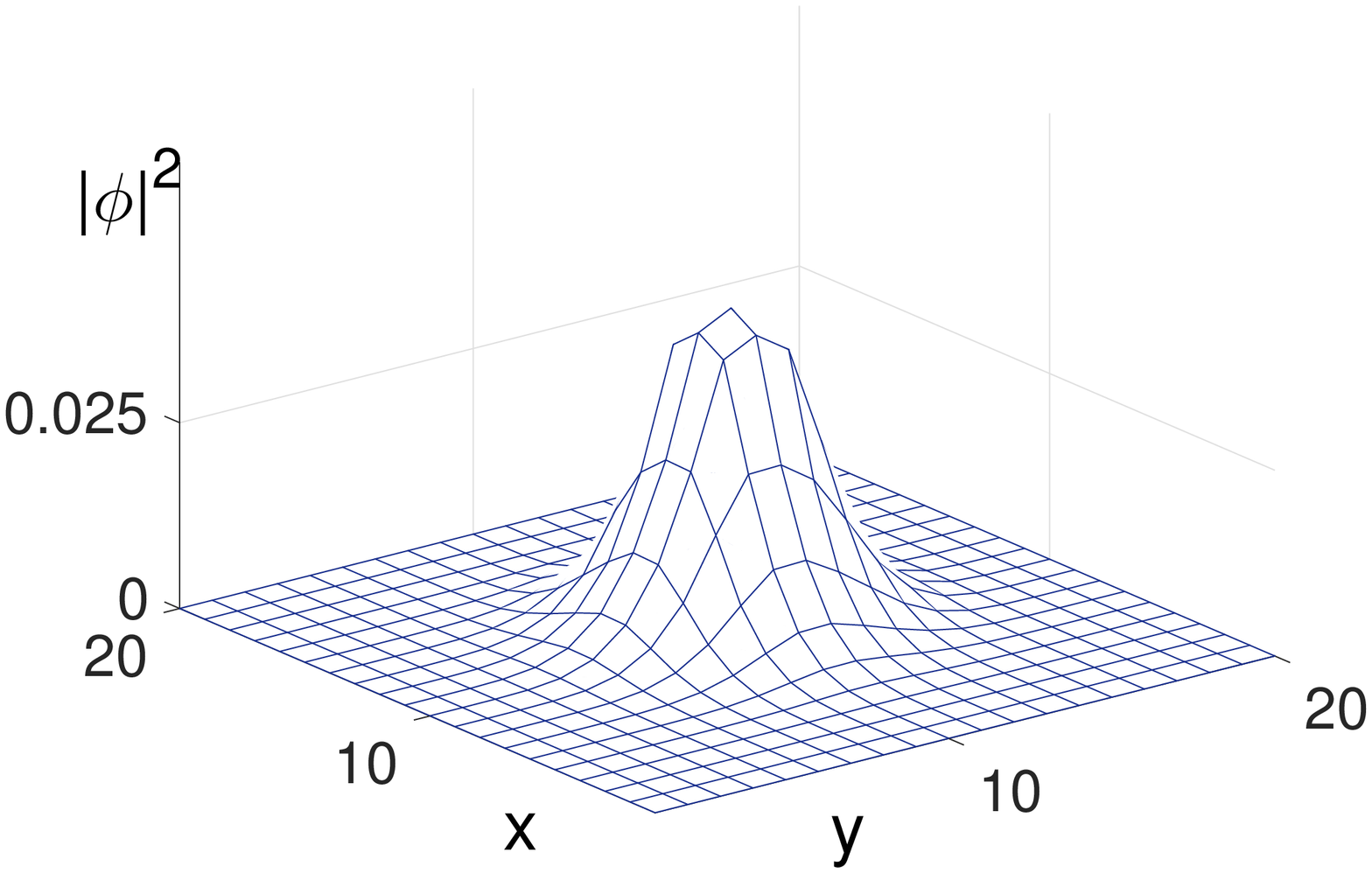}}
\caption{ (a) Quasienergy bands $\varepsilon(k_z)$ of an open-boundary sample with a Floquet line defect (with $n=1$). The system size is $L_x\times L_y\times L_z=20\times 20\times\infty$. The thick blue curve represents the chiral modes localized near $r=0$, while the thin green curve represents the back-propagating modes at the system boundary. Each band is doubly degenerate. (b) The wavefunction profiles of the two energetically degenerate chiral modes at $k_z=0.1\pi$ (indicated by a hollow circle in (a)). (c) shows bands for the $n=2$ defect. The chiral-mode profiles at $k_z=0.3\pi$ and $-0.3\pi$ are shown in (d) and (e), respectively.  }  \label{chiral}
\end{figure}

We calculated the quasienergy bands for a sample with open-boundary in the $x$-$y$ directions, with a defect at the center. In the calculation, $H_0(\bk)$ is Fourier-transformed to the real space ($H_d$ contains $\theta$ but not $\bk$, thus it is already a real-space expression).  For $n=1$, the quasienergy bands is shown in Fig.\ref{chiral}a, in which two in-gap chiral modes with degenerate quasienergy are found (the inessential twofold degeneracy can be lifted by breaking crystal symmetries). The wavefunction profiles indicate that they are sharply localized around the line defect at $r=0$ (Fig.\ref{chiral}b).

For the $n=2$ defect, we find four chiral modes (two thick blue curves in Fig.\ref{chiral}c, each being doubly degenerate) propagating in the same direction as in $n=1$.  In addition to these chiral modes, there are several trivial non-chiral defect modes, which are shown in dashed curves. The profiles of the four chiral modes are shown in Fig.\ref{chiral}d and Fig.\ref{chiral}e.
Summarizing the results for $n=1,2$ and other $n$'s we calculated, the number of chiral modes in the $\omega/2$ bulk gap is \bea M=-2n, \label{N} \eea where the $\pm$ sign stands for $\pm z$ direction of propagation. The factor ``2'' here is somewhat unexpected. Recall that when Dirac fermions are coupled to a complex-valued scalar field\cite{callan1985,witten1985superconducting,jackiw1981}, which serves as a Dirac mass, chiral-mode number of a line defect equals the winding number of complex scalar field, without factor of 2. Our Floquet model differs in that the resonance surface is two-dimensional, thus the defect here belongs to a novel class, to which our intuition from gapping out zero-dimensional Dirac points is inapplicable.

Eq.(\ref{N}) can be predicted by the topological invariant Eq.(\ref{W}). The calculation of $W(\omega/2)$ simplifies significantly in the small $D$ regime. It is sufficient to focus on this regime because $W$, as an integer by definition, is insensitive to the value of $D$, moreover, $D$ is indeed small in current experimental setups\footnote{Calculation for general $D$ is in principle possible by discretizing the Brillouin zone \cite{fukui2005chern,wang2006ab,wu2013bloch}, though more time-consuming. We shall not pursue it here because the small-$D$ calculation suffices for our purpose.}. When $D\ll 1$, $H_d(t)$'s contribution to the integral in Eq.(\ref{W}) is negligible in most region of the $(\bk,\theta,t)$ space, except in small neighborhoods of singular points, where $\frac{\partial U_{\omega/2}}{\partial D}|_{D=0}$ diverges, so that even an infinitesimal $D$ can have non-negligible contribution to the integral. From Eq.(\ref{U-epsilon}), we see that such a divergence can originate only from the branch cut in the definition of $H^{\rm eff}_{\omega/2}(\bk,\theta)$. To obtain $H^{\rm eff}_{\omega/2}$, we recast the full-period evolution operator into\cite{supplemental} \bea U(\bk,\theta,T)= R(T)\,\mathcal{T}\exp\left(-i\int_0^T dt R^\dag(t)[H(t)-i\partial_t]R(t)\right) R^\dag(0), \nn \eea which is valid for any choice of unitary matrix $R(t)$.
Motivated by the rotating-wave approximation\cite{lindner2011floquet,Lindner2013}, we take $R(t)=\exp[\frac{i\omega t}{2}(I-\hat{\bd}\cdot{\bf \Gamma})]$ ($\hat{\bd}=\bd/d$), which satisfies $R(T)=R(0)=I$. Taking its logarithm, we can obtain a formula for $H^{\rm eff}_{\omega/2}$\cite{supplemental} \bea  H^{\rm eff}_{\omega/2}=-\frac{\omega}{2}\hat{\bd}_{\rm R}\cdot{\bf \Gamma}+\cdots, \eea which is accurate to the leading order of $D$. Here, $\hat{\bd}_{\rm R}=\bd_{\rm R}/|\bd_{\rm R}|$ and \bea \bd_{\rm R}\equiv (d-\frac{\omega}{2})\hat{\bd}+\bar{\bD}, \eea with $\bar{\bD}\equiv\bD-(\bD\cdot\hat{\bd})\hat{\bd}$ being the perpendicular (to $\bd$) part of $\bD=(0,0,0,D\cos n\theta, D\sin n\theta)$.
The subscript ``R'' indicates its close relation to the rotating-wave approximiation\cite{lindner2011floquet,Lindner2013}.

Let us write $U_{\omega/2}(t)=\tilde{U}(t)\exp[-i\omega t(I+\hat{\bd}_{\rm R}\cdot{\bf \Gamma})/2]$, then its singular behavior in the $D\rw 0$ limit comes solely from the $\hat{\bd}_{\rm R}\cdot{\bf \Gamma}$ term, and $\tilde{U}(t)$ is nonsingular as $D\rw 0$. Thus we may simply let $D=0$ in $\tilde{U}(t)$, and
the calculation of topological invariant becomes mathematically equivalent to that of a static Hamiltonian $(I+\hat{\bd}_{\rm R}\cdot{\bf \Gamma})/2$, and a straightforward calculation leads to\cite{supplemental}
\bea W(\omega/2)= \frac{3}{8\pi^2}\int d\theta d^3k \epsilon^{\mu\nu\rho\sigma\tau} \frac{d_{{\rm R}\mu}}{d_{\rm R}^5} \frac{\partial d_{{\rm R}\nu}}{\partial k_x}\frac{\partial d_{{\rm R}\rho}}{\partial k_y}\frac{\partial d_{{\rm R}\sigma}}{\partial k_z}\frac{\partial d_{{\rm R}\tau}}{\partial\theta},\quad\quad \eea which can be calculated numerically\footnote{A shortcut\cite{wang2010a} that leads to the same result is to count the number of inverse images of any given point on the 4D unit sphere.}. It is found that\cite{supplemental}  \bea W(\omega/2)=-2n, \eea which precisely matches the number of modes, Eq.(\ref{N}).

By the same calculation, we can also obtain that $W(-\omega/2)=-2n$, therefore,
the second Chern number $C_2(-\omega/2,\omega/2)=W(\omega/2)-W(-\omega/2)=0$.
Thus the Floquet chiral modes in our model have no static analogue, i.e. they are \emph{anomalous} in the terminology of Ref.\cite{rudner2013anomalous,Leykam2016,Mukherjee2016}. In static cases, the chiral-mode number is the sum of all the second Chern numbers of occupied bands\cite{teo2010}, consequently, chiral mode is absent when every Chern number vanishes.

\begin{figure}
\subfigure{\includegraphics[width=8cm, height=4cm]{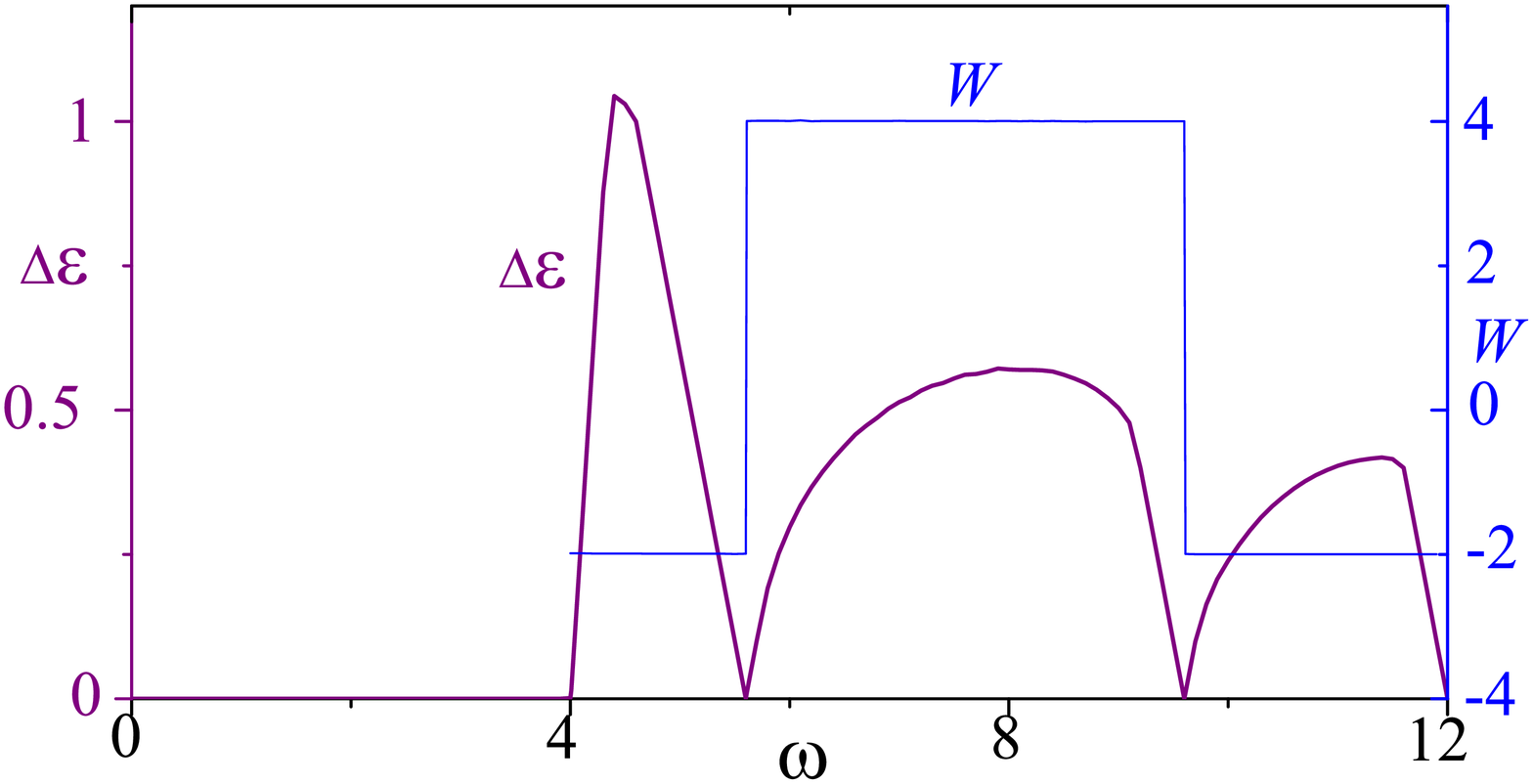}}
\caption{ The bulk quasienergy gap $\Delta\varepsilon$ around $\omega/2$ and the topological invariant $W(\omega/2)$ for $n=1$ (in the gapped regime). }  \label{omega}
\end{figure}

For completeness, we plot the topological invariant as a function of $\omega$ (Fig.\ref{omega}). The number of chiral modes is consistent with its prediction\cite{supplemental}. The topological invariant is not definable for $\omega<4.0$, because the quasienergy gap closes around $k_z=\pi$, which is due to the invasion of the $m=-1$ and $m=2$ Floquet bands into the $\omega/2$ gap. Nevertheless, the chiral modes near $k_z=0$ persist\cite{supplemental}, for the reason that the chiral modes at $\omega/2$ come from the $m=0$ and $m=1$ bands. Without the protection of bulk quasienergy gap, the chiral modes of $\omega<4.0$ can leak into the bulk sample.

{\it Experimental estimations.--}For a typical Dirac semimetal, we estimated that the needed laser frequency is in the visible light regime\cite{supplemental}.  The penetration depth of laser into the sample is estimated to be several hundreds of unit cells\cite{supplemental}. If a film with such thickness is grown, suitable lasers can generate Floquet chiral channels bridging the top and bottom surfaces, which can be measured in transport.

{\it Conclusions.--}We investigated the possibility of creating chiral modes in 3D materials without static topological defect, by exerting a periodic driving with spatial modulation. We  demonstrated it in a concrete model system, moreover, we defined a precise topological invariant, which has the novel feature of combining three classes of variables: momentum, space, and time. Hopefully this work will stimulate further investigations into Floquet topological defects.

Experimentally, this proposal may be realized in Dirac semimetals with appropriate light-matter interaction, as discussed above. If realized, the optically-controllable chiral channels may be useful in future high-speed electronics. Our proposal may also be realized in shaking optical lattices\cite{jotzu2014experimental,parker2013direct,Hauke2012, Zheng2014,Mei2014,Jimenez2015,flaschner2016experimental} with suitable spatial modulation, and phononic (or acoustic) systems\cite{susstrunk2015observation,Prodan2009phonon, wang2015phononic,peano2015topological, mousavi2015topologically,nash2015topological, yang2015topological,kane2014topological,li2011tunable, paulose2015topological,Chen2016tunable, Fleury2016,peng2016experimental}, where the mechanical driving can be made as vortex-shaped by design.

\emph{Acknowledgements.--} R.B., Z.Y., and Z.W. were supported by NSFC (No. 11674189). Z.Y. was supported in part by China Postdoctoral Science
Foundation (No. 2016M590082). L.L was supported by the Ministry of Science and Technology of China (No.
2016YFA0302400) and the National Thousand-Young-Talents Program of China.

\bibliography{dirac}

\vspace{8mm}

{\bf Supplemental Material}

\vspace{4mm}

\section{I. Time-independent limit of the topological invariant}

One of the tests of the validity of the topological invariant $W(\varepsilon)$ is that, in the static limit, $W(\varepsilon)$ should reduce to the static topological invariant. We will show in this limit that $W(\varepsilon)$ does reduce to the second Chern number, which is known to be the correct topological invariant for static line defects.

For the sake of simplicity, we carry out the calculation for flat-band models. ( Non-flat bands can always be smoothly deformed to flat-bands, without changing the topological invariant ).
Let $E_1$ be the energy level of the valence bands (occupied bands), and $E_2$ be the energy level of the conduction band (empty bands).
The flat-band static Hamiltonian can be written as \bea H_0(\bk,\theta)= E_1 P(\bk,\theta) + E_2 [1-P(\bk,\theta)], \eea where $E_1<E_2$ are two constants with the dimension of energy, and $P(\bk)=\sum_{n\in{\rm occ}}|\bk,\theta,n\ra\la\bk,\theta,n|$ is the occupied-bands projection operator, $\{|\bk,\theta,n\ra\}$ being an orthonormal basis of the occupied bands. The projection operators apparently satisfy $P^2(\bk,\theta)=P(\bk,\theta)$ and $(1-P(\bk,\theta))^2=1-P(\bk,\theta)$.

Since we are considering the static limit, i.e. no driving, we can freely choose the driving frequency $\omega$ in the calculation of topological invariant $W(\varepsilon)$. (Adding a zero-amplitude driving with an arbitrary frequency amounts to doing nothing.)  Hereafter we define $\omega=E_2-E_1$, more specifically, $E_1=-\omega$ and $E_2=0$. The advantage of this choice is that $U(\bk,\theta,T)=1$.

In this static limit, the evolution operator becomes: \bea U(\bk,\theta,t)= (e^{i\omega t}-1)P(\bk,\theta)+1, \label{Ustatic} \eea and its inverse is \bea U^{-1}(\bk,\theta,t)=(e^{-i\omega t}-1)P(\bk,\theta)+1. \eea
For notational simplicity, hereafter we introduce $k_{1,2,3}=k_{x,y,z},\,k_4=\theta$, and $\partial_i=\partial/\partial k_i$ (Remark: 4D Floquet topological insulators can also be described by $W(\varepsilon)$ if $k_4$ is regarded as a momentum variable instead of the polar angle $\theta$). By straightforward calculations, we have \bea U^{-1}\partial_t U=i\omega P,  \eea
\bea U^{-1}\partial_i U &=& 2[1-\cos(\omega t)]P\partial_i P + (e^{i\omega t}-1)\partial_i P \nn \\ &=& u(t) P\partial_i P + v(t)\partial_i P,  \eea where we have defined \bea u(t)=2[1-\cos(\omega t)],\,\,\, v(t)=e^{i\omega t}-1. \eea Because the effective Hamiltonian $H^{\rm eff}_\varepsilon=0$ for the flat-band models, we have $U_\varepsilon(k,t)=U(k,t)$.
Now the topological invariant can be simplified to (for any energy $\varepsilon$ other than multiples of $\omega$) \bea W(\varepsilon) &=&\frac{i}{480\pi^3}\int dtd^4k \,{\rm Tr} [\epsilon^{\mu\nu\rho\sigma\tau} (U_\varepsilon^{-1}\partial_\mu U_\varepsilon )( U_\varepsilon^{-1}\partial_\nu U_\varepsilon ) \nn \\ && \times ( U_\varepsilon^{-1}\partial_\rho U_\varepsilon ) (U_\varepsilon^{-1}\partial_\sigma U_\varepsilon)(   U_\varepsilon^{-1}\partial_\tau U_\varepsilon)] \nn \\
&=& \frac{-\omega}{96\pi^3}\int dt d^4k\,{\rm Tr}[\epsilon^{ijkl}P(uP\partial_i P+v\partial_iP)(uP\partial_j P+v\partial_jP)\nn\\&&\times(uP\partial_kP+v\partial_kP)(uP\partial_l P+v\partial_lP)] \nn \\
&=& \frac{-\omega}{96\pi^3}\int dt d^4k\,{\rm Tr}[\epsilon^{ijkl} (u+v)P\partial_i P(uP\partial_j P+v\partial_jP)\nn\\&&\times(uP\partial_kP+v\partial_kP)(uP\partial_l P+v\partial_lP)]. \label{S1} \eea
To simplify this expression, we notice that \bea P(\partial_iP)P=P\partial_i(PP)-PP\partial_iP =0, \eea then Eq.(\ref{S1}) can be reduced to \bea W(\varepsilon)
&=& \frac{-\omega}{96\pi^3}\int dt d^4k\,{\rm Tr}[\epsilon^{ijkl} (u+v)v P\partial_i  P\partial_jP \nn\\ && \times(uP\partial_kP+v\partial_kP)(uP\partial_l P+v\partial_lP)] \nn\\&=& \frac{-\omega}{96\pi^3}\int dt d^4k\,{\rm Tr}[\epsilon^{ijkl} (u+v)v\nn\\ &&(uv P\partial_i P\partial_jP P\partial_kP\partial_lP + v^2P\partial_iP\partial_jP\partial_kP\partial_lP)]. \label{S2} \eea
The two terms here are in fact proportional. Indeed, by an integration by parts, we have \bea  &&\int d^4k\,{\rm Tr}[\epsilon^{ijkl}P\partial_i P\partial_jP P\partial_kP\partial_lP] \nn\\ &=& \int d^4k\,{\rm Tr}\left(\epsilon^{ijkl} P\partial_iP[\partial_j(PP)-P\partial_jP]\partial_kP\partial_lP\right) \nn\\ &=& \int d^4k\,{\rm Tr}[\epsilon^{ijkl} P\partial_iP \partial_jP \partial_kP\partial_lP], \label{equal-integral}\eea where we have used the identity $P(\partial_iP)P=0$ to get the last line.

With Eq.(\ref{equal-integral}) as an input, Eq.(\ref{S2}) becomes \bea W(\varepsilon)
&=&\frac{-\omega}{96\pi^3}\int dt d^4k\,{\rm Tr}[\epsilon^{ijkl} (u+v)^2v^2 \nn\\ && P\partial_i P\partial_jP P\partial_kP\partial_lP]. \label{S3} \eea The next step is to integrate out $t$. By straightforward calculation, we can see that \bea \int_0^{2\pi/\omega}dt\,[u(t)+v(t)]^2 v^2(t) = 12\pi/\omega, \label{integral} \eea thus we finally have \bea W(\varepsilon)
=-\frac{1}{8\pi^2}\int d^4k\,{\rm Tr}[\epsilon^{ijkl} P\partial_i P\partial_jP P\partial_kP\partial_lP],  \eea in which the $\omega$ prefactor in Eq.(\ref{S3}) has been neatly canceled by the $1/\omega$ factor in Eq.(\ref{integral}). This is the second Chern number of all the occupied bands, expressed in terms of the projection operator, of a time-independent Hamiltonian\cite{qi2008} (But remember that $k_4=\theta$ here). Therefore, the topological invariant $W(\varepsilon)$ reduces to the second Chern number in the static limit. Thus the static topological invariant is recovered as a special case (when the driving vanishes) of $W(\varepsilon)$.


\section{II. Proof of $W(\varepsilon')-W(\varepsilon)=C_2(\varepsilon,\varepsilon')$}

First of all, from the definition of $U_{\varepsilon}$, we can see  that (see also Refs.\cite{rudner2013anomalous,Carpentier2015})
\bea U^{-1}_{\varepsilon}U_{\varepsilon'}=\exp(i\omega t P_{\varepsilon,\varepsilon'}), \label{minus} \eea
thus we can define $\bar{U}_{\varepsilon,\varepsilon'}(\bk,\theta,t)=\exp(i\omega t P_{\varepsilon,\varepsilon'})$, and Eq.(\ref{minus}) tells us that \bea W(\varepsilon')-W(\varepsilon)=W(\bar{U}_{\varepsilon,\varepsilon'}),\eea where $W(\bar{U}_{\varepsilon,\varepsilon'})$ is defined as replacing $U_\varepsilon$ by $\bar{U}_{\varepsilon,\varepsilon'}$ in the definition of $W$. An equivalent expression for $\bar{U}_{\varepsilon,\varepsilon'}$ is
\bea \bar{U}_{\varepsilon,\varepsilon'}(\bk,\theta,t)= (e^{i\omega t}-1)P_{\varepsilon,\varepsilon'}(\bk,\theta)+1, \eea which takes the same form as Eq.(\ref{Ustatic}), therefore, the same calculations as the previous section lead to \bea W(\bar{U}_{\varepsilon,\varepsilon'}) = C_2(\varepsilon,\varepsilon'), \eea which finishes the proof of $W(\varepsilon')-W(\varepsilon)=C_2(\varepsilon,\varepsilon')$.

\section{III. Hamiltonian in the real space for a homogeneous system}

In the main article, we have defined five $\Gamma$ matrices:
\bea
\Gamma_1&=&\sigma_1\otimes\tau_3=\begin{pmatrix}
\sigma_1 & 0 \\
0 & -\sigma_1
\end{pmatrix}, \nn \\
\Gamma_2&=&\sigma_2\otimes\tau_3=\begin{pmatrix}
\sigma_2 & 0 \\
0 & -\sigma_2
\end{pmatrix}, \nn \\
\Gamma_3&=&\sigma_3\otimes\tau_3=\begin{pmatrix}
\sigma_3 & 0 \\
0 & -\sigma_3
\end{pmatrix},\nn \\
\Gamma_4&=&\sigma_0\otimes\tau_1=\begin{pmatrix}
0 & \sigma_0 \\
\sigma_0 & 0
\end{pmatrix},\nn \\
\Gamma_5&=&\sigma_0\otimes\tau_2=\begin{pmatrix}
0 & -i\sigma_0 \\
i\sigma_0 & 0
\end{pmatrix},
\eea
The $\bk$-space Hamiltonian for a homogeneous system reads
\begin{equation}
H(\bk,t)=H_0(\bk)+H_d(t),
\end{equation}
where the explicit expressions of $H_0(\bk)$ and $H_d(t)$ are
\begin{align}
H_0(\bk)=&2t_x\sin k_x\Gamma_1+2t_y\sin k_y\Gamma_2+2t_z\sin k_z\Gamma_3 \nonumber\\
+&(m-B_1\sum_{i=x,y,z}\cos k_i-B_2\sum_{i\neq j}\cos k_i\cos k_j)\Gamma_4,\\
H_d(t)=&2D\cos(\omega t)(\cos \alpha\Gamma_4+\sin \alpha\Gamma_5),
\end{align}
For the sample that is finite in the $x$ and $y$ direction, and infinite in the $z$ direction, we may keep $k_z$, and do a Fourier transformation in the $x$ and $y$ directions. The resultant Hamiltonian is
\begin{equation}
\hat{H}(k_z,t)=\sum_{x,y;x',y'}c^{\dagger}_{x,y}H_{x,y;x',y'}(k_z,t)c_{x',y'}+h.c.,
\end{equation}
where $c^{\dagger}_{x,y}$ and $c_{x',y'}$ are fermion operators ($k_z$ is implicit). The nonvanishing elements of $H_{x,y;x',y'}(t)$ in our model are
\begin{align}
H_{x+1,y;x,y}&=it_x\Gamma_1-(\frac{B_1}{2}+B_2\cos k_z)\Gamma_4, \\
H_{x,y;x+1,y}&=-it_x\Gamma_1-(\frac{B_1}{2}+B_2\cos k_z)\Gamma_4, \\
H_{x,y+1;x,y}&=it_y\Gamma_2-(\frac{B_1}{2}+B_2\cos k_z)\Gamma_4, \\
H_{x,y;x,y+1}&=-it_y\Gamma_2-(\frac{B_1}{2}+B_2\cos k_z)\Gamma_4, \\
H_{x+1,y+1;x,y}&=H_{x,y;x+1,y+1}=-\frac{B_2}{2}\Gamma_4, \\
H_{x+1,y;x,y+1}&=H_{x,y+1;x+1,y}=-\frac{B_2}{2}\Gamma_4, \\
H_{x,y;x,y} &=2t_z\sin k_z\Gamma_3+(m-B_1\cos k_z)\Gamma_4 \nonumber\\
&+2D\cos(\omega t)(\cos \alpha\Gamma_4+\sin \alpha\Gamma_5).
\end{align}

\section{IV. Hamiltonian in the real space for a Floquet defect}

In the presence of a Floquet topological defect (see the main article), the real-space Hamiltonian is almost the same as that of the homogeneous system, as given in the previous section, except that the spatially uniform parameter $\alpha$ should now be taken as $\alpha=n\theta$ ($n$ is a nonzero integer), where $\theta$ is the polar angle, namely,
\begin{equation}
\theta_{x,y}=\text{arctan}\frac{y-y_0}{x-x_0}.
\end{equation}
The real-space Hamiltonian is
\begin{equation}
\hat{H}(k_z,t)=\sum_{x,y;x',y'}c^{\dagger}_{x,y}H_{x,y;x',y'}(k_z,t)c_{x',y'}+h.c.,
\end{equation}
in which the nonvanishing elements of $H_{x,y;x',y'}$ in our model are
\begin{align}
H_{x+1,y;x,y}&=it_x\Gamma_1-(\frac{B_1}{2}+B_2\cos k_z)\Gamma_4, \\
H_{x,y;x+1,y}&=-it_x\Gamma_1-(\frac{B_1}{2}+B_2\cos k_z)\Gamma_4, \\
H_{x,y+1;x,y}&=it_y\Gamma_2-(\frac{B_1}{2}+B_2\cos k_z)\Gamma_4, \\
H_{x,y;x,y+1}&=-it_y\Gamma_2-(\frac{B_1}{2}+B_2\cos k_z)\Gamma_4, \\
H_{x+1,y+1;x,y}&=H_{x,y;x+1,y+1}=-\frac{B_2}{2}\Gamma_4, \\
H_{x+1,y;x,y+1}&=H_{x,y+1;x+1,y}=-\frac{B_2}{2}\Gamma_4, \\
H_{x,y;x,y} &=2t_z\sin k_z\Gamma_3+(m-B_1\cos k_z)\Gamma_4 \nonumber\\
&+2D\cos(\omega t)(\cos n\theta_{x,y}\Gamma_4+\sin  n\theta_{x,y}\Gamma_5). \label{defect-element}
\end{align}
Compared to the real-space Hamiltonian of a spatially uniform system (see the previous section), the only modification is in $H_{x,y;x,y}$ (Eq.\ref{defect-element}), which now contains the polar angle $\theta_{x,y}$.

\section{V. Technical details in the calculation of the topological invariant $W(\omega/2)$}

Let us first derive the equation $U(\bk,\theta,T)=R(T)\exp\left(-i\int_0^T dt R^\dag(t)[H(t)-i\partial_t]R(t)\right)R^\dag(0)$, which is useful in the main article.  To this end, let us express the integral $U(\bk,\theta,t)= \mathcal{T}\exp\left(-i\int_0^T dt  H(\bk,\theta,t)\right)$ as product after discretizing the interval $[0,T]$ as $\{ 0,\tau,2\tau,3\tau,\cdots,T\}$, $T/\tau$ being a large integer, then we have
\begin{widetext}
\bea U(\bk,\theta,T)&=& \prod_j\cdots\exp[-i\tau H((j+1)\tau)]\exp[-i\tau H(j\tau)]\exp[-i\tau H((j-1)\tau)]\cdots,\nn\\ &=& \prod_j\cdots\exp[-i\tau H((j+1)\tau)]R((j+1)\tau)R^\dag((j+1)\tau) \exp[-i\tau H(j\tau)] R(j\tau)R^\dag(j\tau)\exp[-i\tau H((j-1)\tau)]\cdots, \eea where we have kept implicit the $(\bk,\theta)$ arguments at this stage for notational simplicity.
Making use of \bea R^\dag((j+1)\tau) \exp[-i\tau H(j\tau)] R(j\tau)\approx \exp\left(-i\tau R^\dag(j\tau)[H(j\tau)-i\partial_t]R(j\tau)\right),   \eea where ``$\approx$'' becomes ``$=$'' in the $\tau\rw 0$ limit, we obtain \bea U(\bk,\theta,T)&=&  \prod_j \cdots \exp\left(-i\tau R^\dag((j+1)\tau)[H((j+1)\tau)-i\partial_t]R((j+1)\tau)\right) \exp\left(-i\tau R^\dag(j\tau)[H(j\tau)-i\partial_t]R(j\tau)\right)\cdots \nn\\ &=& R(T)\,\mathcal{T}\exp\left(-i\int_0^T dt \,R^\dag(t)[H(t)-i\partial_t]R(t)\right)\,R^\dag(0), \eea
\end{widetext}
We take \bea R(t) =
\exp[\frac{i\omega t}{2}(I-\hat{\bd}\cdot{\bf \Gamma})], \eea where $I$ stands for the identity matrix. This choice of $R(t)$ is essentially the rotating-wave approximation\cite{lindner2011floquet}. We can also notice that $R(T)=R(0)=I$. Now we have \bea  && R^\dag(t)[H(t)-i\partial_t]R(t) \nn\\ &=& H_0+R^\dag(t)H_d(t)R(t)-\frac{\omega}{2}\hat{\bd}\cdot{\bf \Gamma} +\frac{\omega}{2} \nn \\ &=& H_0 + (\bar{\bD}\cdot{\bf \Gamma} + \cdots )-\frac{\omega}{2}\hat{\bd}\cdot{\bf \Gamma} +\frac{\omega}{2},  \eea where $\bar{\bD}=\bD-(\bD\cdot\hat{\bd})\hat{\bd}$ with $\bD=(0,0,0,D\cos n\theta, D\sin n\theta)$, and ``$\cdots$'' in the parentheses stands for terms proportional to $\exp(\pm i\omega t)$. To the leading order of $D$, we have  \bea U(\bk,\theta,T) &=&  \mathcal{T}\exp\left(-i\int_0^T dt \,R^\dag(t)[H(t)-i\partial_t]R(t)\right)\nn\\ &=&  \exp[-iT(\bd_{\rm R}\cdot{\bf \Gamma}+\frac{\omega}{2})], \label{U-1} \eea in which $\bd_{\rm R}\equiv (d-\frac{\omega}{2})\hat{\bd}+\bar{\bD}$. The $\exp(\pm i\omega t)$ terms in $R^\dag(t)[H(t)-i\partial_t]R(t)$ do not contribute at the leading order of $D$. Near the resonance surface $d(\bk)=\omega/2$, we have $|\bd_{\rm R}|\ll 1$, yet the $\bd_{\rm R}\cdot{\bf \Gamma}$ term is non-negligible due to the branch cut in the definition of $H^{\rm eff}_{\omega/2}$. The branch cut generates a $\frac{\hat{\bd}_{\rm R}\cdot{\bf \Gamma}+1}{2}$ term. Therefore, to the leading order of $D$, we have \bea H^{\rm eff}_{\omega/2} &=& \frac{\omega}{2}-\omega\frac{\hat{\bd}_{\rm R}\cdot{\bf \Gamma}+1}{2}+\cdots\nn\\ &=& -\frac{\omega}{2}\hat{\bd}_{\rm R}\cdot{\bf \Gamma}+\cdots  \eea near the resonance surface. We can see that $\frac{\partial H^{\rm eff}_{\omega/2}}{\partial D}|_{D=0}$ diverges at the resonance surface $d(\bk)=\omega/2$, which means that an infinitesimal $D$ has non-negligible contribution to the integral in the definition of $W_{\omega/2}$ (Eq.4 in the main article). It follows that the integral receives contribution mainly from the neighborhood of the resonance surface when $D\ll 1$.

The topological invariant $W(\omega/2)$ is defined in terms of $U_{\omega/2}(t)$, which is given by \bea U_{\omega/2}(t)=U(t)\exp(iH^{\rm eff}_{\omega/2}t). \eea
There is a simple trick to avoid coping with $U(t)$ in this formula. We consider the difference between the topological invariant $W(\omega/2)$ for two values of $n$, denoted as $n_1$ and $n_2$. For clarity, they are denoted as $W(\omega/2,n_1)$ and $W(\omega/2,n_2)$. The trick is to calculate $W(\omega/2,n_1)-W(\omega/2,n_2)$, which is given by the integration in Eq.(4), with $U_{\varepsilon}$ replaced by $U_{\omega/2}(t,n_1)U^{-1}_{\omega/2}(t,n_2)$ (Hereafter, this integral is denoted as $W[U^{-1}_{\omega/2}(t,n_2)U_{\omega/2}(t,n_1)]$). In fact, we have \bea W[U^{-1}_{\omega/2}(t,n_2)U_{\omega/2}(t,n_1)] &=& W[U_{\omega/2}(t,n_1)]+W[U^{-1}_{\omega/2}(t,n_2)] \nn \\
&=& W[U_{\omega/2}(t,n_1)]-W[U_{\omega/2}(t,n_2)]\nn \\
&\equiv& W(\omega/2,n_1) - W(\omega/2,n_2), \eea due to the additive property of the winding number. Taking $n_1=n,n_2=0$ leads to \bea W[U^{-1}_{\omega/2}(t,0)U_{\omega/2}(t,n)] = W(\omega/2,n), \eea where we have used the apparent fact that $W(\omega/2,0)=0$ (when $U$ is independent of $\theta$, $W$ must vanish).

The trick is useful because the calculation of $W[U^{-1}_{\omega/2}(t,0)U_{\omega/2}(t,n)]$ is easier than that of $W[U_{\omega/2}(n)]$. In fact, we have \bea && U^{-1}_{\omega/2}(t,0)U_{\omega/2}(t,n)\nn\\ &=& \exp[-iH^{\rm eff}_{\omega/2}(0)t]U^{-1}(t,0)U(t,n)\exp[iH^{\rm eff}_{\omega/2}(n)t]\nn\\ &\approx& \exp[-iH^{\rm eff}_{\omega/2}(0)t]\exp[iH^{\rm eff}_{\omega/2}(n)t]\nn\\ &\approx& \exp[i\omega\frac{\hat{\bd}_{\rm R}(0)\cdot{\bf \Gamma}+1}{2}t]\exp[-i\omega\frac{\hat{\bd}_{\rm R}(n)\cdot{\bf \Gamma}+1}{2}t].  \eea  The above simplification of eliminating $U(t)$ occurs because $U^{-1}(t,0)U(t,n)\approx I$ (identity matrix) when $D\ll 1$. Now the calculation of $W[U^{-1}_{\omega/2}(t,0)U_{\omega/2}(t,n)]$ is simplified to be mathematically equivalent to the case of static Hamiltonian. By a straightforward calculation that is essentially the same as Sec.I in this Supplemental Material, we have
\bea
W(\omega/2,n)= \frac{-1}{8\pi^2}\int d^4k\,{\rm Tr}[\epsilon^{ijkl} P\partial_i P\partial_jP P\partial_kP\partial_lP], \eea where $P\equiv \frac{\hat{\bd}_{\rm R}(n)\cdot{\bf \Gamma}+1}{2}$. By a calculation similar to that in Ref.\cite{qi2008}, it can be simplified to  \bea W(\omega/2,n) &=&\frac{3}{8\pi^2}\int d\theta d^3k \epsilon^{\mu\nu\rho\sigma\tau} \hat{d}_{{\rm R}\mu}\frac{\partial \hat{d}_{{\rm R}\nu}}{\partial k_x}\frac{\partial \hat{d}_{{\rm R}\rho}}{\partial k_y}\frac{\partial \hat{d}_{{\rm R}\sigma}}{\partial k_z}\frac{\partial \hat{d}_{{\rm R}\tau}}{\partial\theta}  \nn \\
&=&\frac{3}{8\pi^2}\int d\theta d^3k \epsilon^{\mu\nu\rho\sigma\tau} \frac{d_{{\rm R}\mu}}{d_{\rm R}^5} \frac{\partial d_{{\rm R}\nu}}{\partial k_x}\frac{\partial d_{{\rm R}\rho}}{\partial k_y}\frac{\partial d_{{\rm R}\sigma}}{\partial k_z}\frac{\partial d_{{\rm R}\tau}}{\partial\theta},\quad\quad\label{W-R}
\eea
where $\mu,\nu,\rho,\sigma,\tau=1,2,3,4,5$. This topological quantity measures the number of times that the unit vector ${\bf \hat{d}}_{\rm R}={\bf  d}_{\rm R}/d_{\rm R}$ covers the 4D unit sphere as $(k_x,k_y,k_z,\theta)$ vary.

\begin{figure}
\includegraphics[width=8cm, height=5cm]{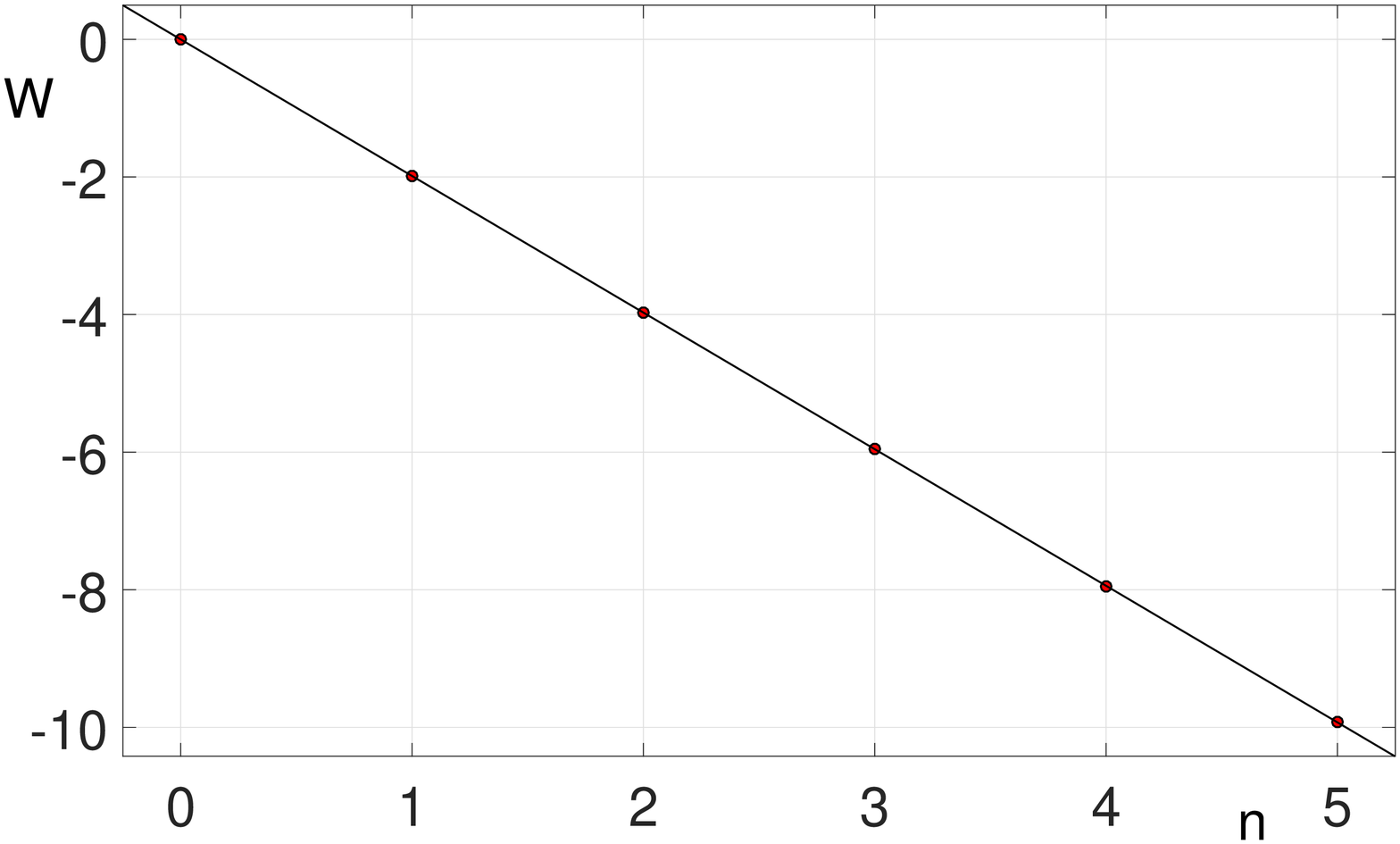}
\caption{Plot of $W(\omega/2)$ as a function of the integer $n$ in numerical integration. We take $\omega=4.2$ and $D=0.2$ in the calculation. The numerical results indicate that $W(\omega/2)=-2n$. }  \label{integration}
\end{figure}

The explicit formulas for all components of $\bd_{\rm R}$ are \bea d_{{\rm R}1}&=& (d-\omega/2)\hat{d}_1+\bar{D}_1,\nn\\ d_{{\rm R}2}&=&(d-\omega/2)\hat{d}_2+\bar{D}_2, \nn\\ d_{{\rm R}3}&=&(d-\omega/2)\hat{d}_3+\bar{D}_3, \nn\\ d_{{\rm R}4}&=& (d-\omega/2)\hat{d}_4+\bar{D}_4,\nn\\d_{{\rm R}5}&=&\bar{D}_5. \eea More explicitly, we have
\bea
d_{\rm{R}1}&=&\hat{d}_1(d-\omega/2-\hat{d}_4 D\cos n\theta), \nn\\
d_{\rm{R}2}&=&\hat{d}_2(d-\omega/2-\hat{d}_4 D\cos n\theta), \nn\\
d_{\rm{R}3}&=&\hat{d}_3(d-\omega/2-\hat{d}_4 D\cos n\theta), \nn\\
d_{{\rm{R}}4}&=&\hat{d}_4(d-\omega/2-\hat{d}_4 D\cos n\theta)+D\cos n\theta, \nn\\
d_{{\rm{R}}5}&=&D\sin n\theta.
\eea
Inserting these expressions into Eq.(\ref{W-R}) and doing a numerical integration, we find that the numerical results (Fig.\ref{integration}) clearly fit
\bea
W(\omega/2,n)=-2n.
\eea

\section{VI. The quasienergy bands and Floquet chiral modes for several other frequencies}

\begin{figure}
\subfigure{\includegraphics[width=9cm, height=5cm]{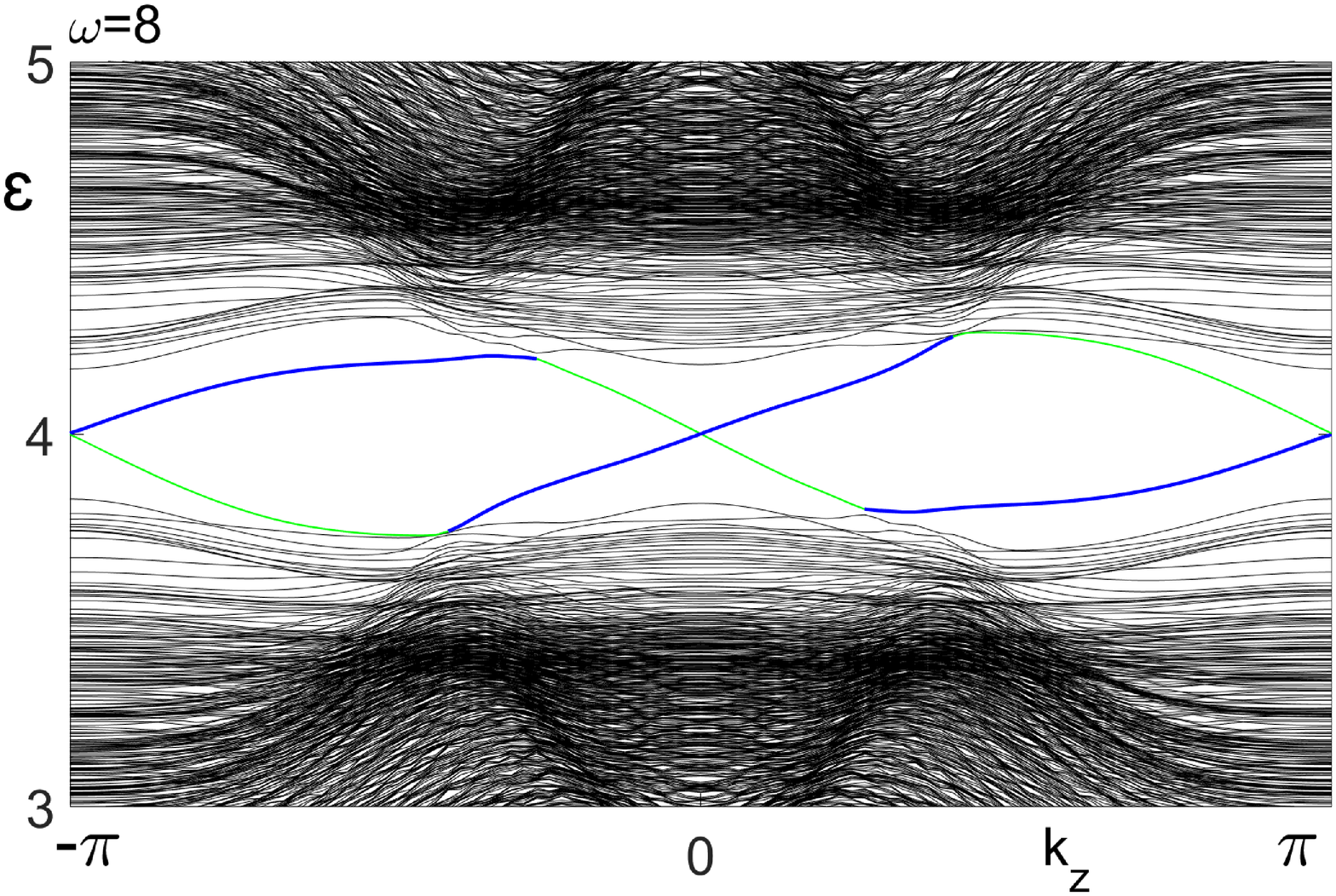}}
\subfigure{\includegraphics[width=9cm, height=5cm]{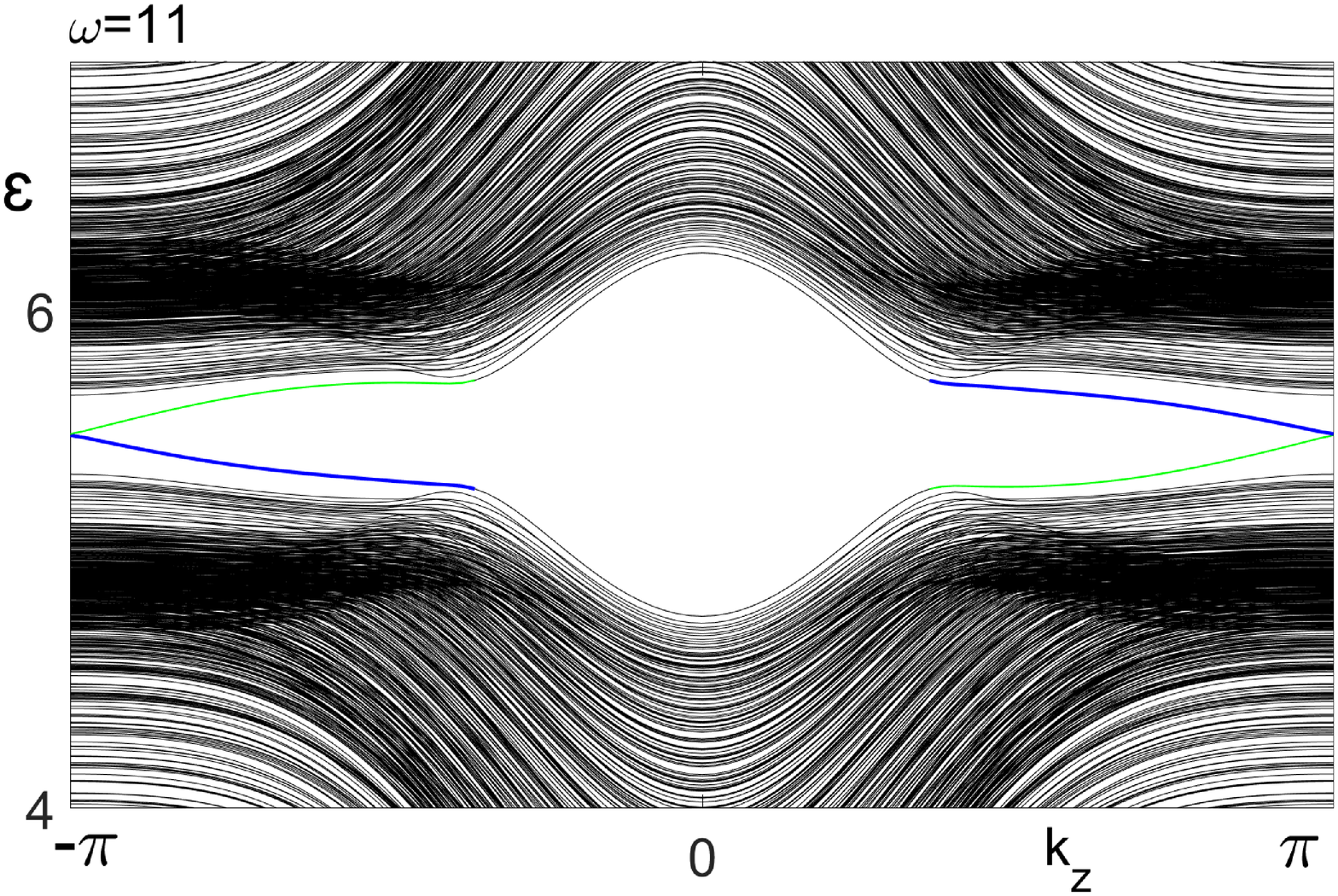}}
\caption{ Quasienergy bands $\varepsilon(k_z)$ of an open-boundary sample with a Floquet line defect (with $n=1$), for $\omega=8.0$ and $\omega=11.0$ (marked in each figure). The system size is $L_x\times L_y\times L_z=20\times 20\times\infty$. The thick blue curve represents the chiral modes localized near $r=0$, while the thin green curve represents the back-propagating modes at the system boundary. Each band is doubly degenerate.  }  \label{8-11}
\end{figure}

In the main article, we have plot the topological invariant $W(\omega/2)$ as a function of $\omega$ (see Fig.4 in the main article). For $\omega=8.0$ and $\omega=11.0$, we have $W(\omega/2)=4$ and $W(\omega/2)=-2$ for the $n=1$ Floquet defect, respectively.

Here, we show the quasienergy bands for $\omega=8.0$ and $\omega=11.0$ (Fig.\ref{8-11}). We can see that the number of the chiral modes and the propagating direction of the modes are consistent with the prediction of topological invariant. We notice that for $\omega=8.0$, there are two chiral modes around $k_z=0$, and two around $k_z=\pi$; for $\omega=11.0$, there are two chiral modes around $k_z=\pi$ (remember the double degeneracy of each band). The topological invariant determines the total number of chiral modes.

\begin{figure}
\subfigure{\includegraphics[width=9cm, height=5cm]{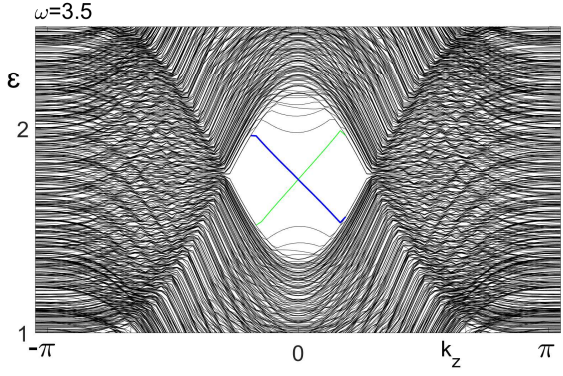}}
\caption{ Quasienergy bands $\varepsilon(k_z)$ of an open-boundary sample with a Floquet line defect (with $n=1$), for $\omega=3.5$. The system size is $L_x\times L_y\times L_z=20\times 20\times\infty$. The thick blue curve represents the chiral modes localized near $r=0$, while the thin green curve represents the back-propagating modes at the system boundary. Each band is doubly degenerate. }  \label{3.5}
\end{figure}

For $\omega=3.5$, there is no quasienergy gap at $\omega/2$, thus the topological invariant cannot be defined. Nevertheless, chiral modes persist, as shown in Fig.\ref{3.5}. Without the protection of the bulk quasienergy gap, these modes can leak into the bulk.

\section{VII. Experimental estimations}

Taking a typical Dirac semimetal metal Cd$_{3}$As$_{2}$ as an example, we will estimate the suitable frequency ($\omega$) of the laser, and the penetration depth ($\delta$) of a laser into the sample.

The penetration depth $\delta(\omega)$ of lasers into the Dirac semimetals can be obtained
by the formula\cite{hummel2011electronic}
\begin{eqnarray}
\delta(\omega)=\frac{n(\omega)\epsilon_{0}c}{\text{Re}\sigma(\omega)},
\end{eqnarray}
where $n(\omega)$  is the refraction index of the materials,
$\epsilon_{0}$ the permittivity of vacuum, $c$ the speed of light in vacuum, and
$\text{Re}\sigma(\omega)$ is the real part of optical conductivity (isotropy is
assumed for simplicity). In the zero temperature and
clean limit (without impurity scattering), under the
neutrality condition (the chemical potential is located at the Weyl point), $\text{Re}\sigma(\omega)$ of a single Weyl cone takes the simple form of \cite{hosur2012,Tabert2016optical}
\begin{eqnarray}
\text{Re}\sigma(\omega)=\frac{e^{2}}{24\pi \hbar v_{F}}\omega.
\end{eqnarray}
A Dirac cone
consists of two Weyl cones of opposite chirality, thus a factor of 2 should be included for a Dirac cone: \begin{eqnarray}
\text{Re}\sigma(\omega)=\frac{e^{2}}{12\pi \hbar v_{F}}\omega.
\end{eqnarray}

In the following, we take the experimentally-confirmed Dirac semimetal Cd$_{3}$As$_{2}$
as a concrete example to estimate the penetration depth. In experiments, it was found that
Cd$_{3}$As$_{2}$ possesses a pair of Dirac cones near the $\Gamma$ point \cite{Liu2014b}.
Thus, if the anisotropy of the Dirac cones is neglected, the real part of
the optical conductivity of Cd$_{3}$As$_{2}$ is approximately given by
\begin{eqnarray}
\text{Re}\sigma(\omega)\approx 2\times\frac{e^{2}}{12\pi \hbar v_{F}}\omega=\frac{e^{2}}{6\pi \hbar v_{F}}\omega,
\end{eqnarray}
where the factor $2$ counts the number of Dirac cones. In the isotropy approximation, we take the Fermi velocity $v_{F}$ as
the average value in three directions, which is\cite{Liu2014b}
\begin{eqnarray}
v_{F}&=&(v_{x}v_{y}v_{z})^{1/3}\nonumber\\
&=&(1.28\times10^{6}\times1.3\times10^{6}\times3.27\times10^{5})^{1/3}\text{m/s}\nonumber\\
&=&8.16\times10^{5}\text{m/s}.
\end{eqnarray}
 Consequently,
\begin{eqnarray}
\delta(\omega)&=&\frac{6\pi \hbar v_{F}n(\omega)\epsilon_{0}c}{e^{2}\omega}\nonumber\\
&=&\frac{3\hbar v_{F}\epsilon_{0}}{e^{2}}n(\omega)\lambda\nonumber\\
&\approx&\frac{3\times1.05\times10^{-34}\times 8.16\times 10^{5}\times8.85\times10^{-12}}{1.6^{2}\times10^{-38}}n(\omega)\lambda\nonumber\\
&\approx&0.089n(\omega)\lambda,
\end{eqnarray}
where $\lambda=2\pi c/\omega$ is the wavelength of the light. To estimate $n(\omega)$ and $\lambda$,
we need an estimation of the bandwidth of the Dirac semimetal.
As a crude estimation using linear dispersion, the bandwidth
is given by
\begin{eqnarray}
E_{\rm bw}= v_{F}\frac{2\pi}{a},
\end{eqnarray}
where $a$ is the lattice constant. For Cd$_{3}$As$_{2}$, the lattice constant between  nearest-neighbour
sites in the natural cleavage plane is $4.6$ {\AA}\cite{Liu2014b}, which leads to
\begin{eqnarray}
E_{\rm bw}&=&8.16\times10^{5}\text{m/s}\times\frac{2\times3.14}{4.6\times10^{-10}\text{m}}\nonumber\\
&\approx&1.11\times10^{16}{\rm s}^{-1} \equiv 7.33 \text{eV}.
\end{eqnarray}
In the main article, the bandwidth of our lattice model is $12.0$ in dimensionless form, while the frequency is taken to be $4.2$. If we replace the bandwidth $12.0$ by $E_{\rm bw}=7.33$eV, then the angular frequency is
\begin{eqnarray}
\omega =\frac{4.2}{12}\times7.33 \text{eV}\approx2.57 \text{eV},
\end{eqnarray}
thus \bea \lambda=2\pi c/\omega \approx484{\rm nm},\eea which is in the visible light regime.
In experiments, many available lasers are in this regime, e.g., He-Cd laser (441.6 nm), Ar laser (488 nm).
At these wavelengthes,  it has been experimentally found that
$n(\omega)\simeq5.2$\cite{Zdanowicz1967Cd3As2}. Therefore, the penetration depth is approximately given by
\begin{eqnarray}
\delta(\omega)&\approx& 0.089\times5.2\times484 \text{nm}\approx224 \text{nm}\approx487a.
\end{eqnarray} Finally, we remark that Cd$_{3}$As$_{2}$ has been taken just for an order of magnitude estimation of penetration depth in Dirac semimetals. To provide the preferred forms of light-matter interaction discussed in the main article, other Dirac semimetals (such as magnetic Dirac semimetals) are likely to be candidates.

\end{document}